\documentclass[aps, pre, twocolumn, a4paper, floatfix, 11point]{revtex4-1}

\usepackage[utf8]{inputenc}
\usepackage{epsfig}
\usepackage[T1]{fontenc}
\usepackage{t1enc}
\usepackage{graphicx}
\usepackage{amssymb}
\usepackage{amsmath}

\newcommand{\avg}[1]{{\left<#1\right>}}

\usepackage{mathtools}
\def\multiset#1#2{\ensuremath{\left(\kern-.3em\left(\genfrac{}{}{0pt}{}{#1}{#2}\right)\kern-.3em\right)}}

\begin{document}

\title{Entropy of stochastic blockmodel ensembles}

\author{Tiago P. Peixoto}
\email{tiago@itp.uni-bremen.de}
\affiliation{Institut f\"{u}r Theoretische Physik, Universit\"at Bremen,
Otto-Hahn-Allee 1, D-28359 Bremen, Germany}


\begin{abstract}
  Stochastic blockmodels are generative network models where the
  vertices are separated into discrete groups, and the probability of an
  edge existing between two vertices is determined solely by their group
  membership. In this paper, we derive expressions for the entropy of
  stochastic blockmodel ensembles. We consider several ensemble
  variants, including the traditional model as well as the newly
  introduced degree-corrected version [Karrer et al. Phys. Rev. E {\bf
  83}, 016107 (2011)], which imposes a degree sequence on the vertices,
  in addition to the block structure. The imposed degree sequence is
  implemented both as ``soft'' constraints, where only the expected
  degrees are imposed, and as ``hard'' constraints, where they are
  required to be the same on all samples of the ensemble. We also
  consider generalizations to multigraphs and directed graphs. We
  illustrate one of many applications of this measure by directly
  deriving a log-likelihood function from the entropy expression, and
  using it to infer latent block structure in observed data. Due to the
  general nature of the ensembles considered, the method works well for
  ensembles with intrinsic degree correlations (i.e. with entropic
  origin) as well as extrinsic degree correlations, which go beyond the
  block structure.
\end{abstract}

\maketitle

\section{Introduction}

Stochastic
blockmodels~\cite{holland_stochastic_1983,fienberg_statistical_1985,
faust_blockmodels:_1992, anderson_building_1992} are random graph
ensembles, in which vertices are separated into discrete groups (or
``blocks''), and the probability of an edge existing between two
vertices is determined according to their group membership. This class
of model (together with many variants which incorporate several other
details~\cite{doreian_generalized_2004,brandes_network_2010}) has been
used extensively in the social sciences, where the blocks usually
represents the roles played by different social agents. In this context,
it has been used mainly as a tool to infer latent structure in empirical
data. More recently, it has been applied as an alternative to the more
specific task of community detection~\cite{fortunato_community_2010},
which focus solely on densely connected communities of
vertices~\cite{newman_mixture_2007, reichardt_role_2007,
bickel_nonparametric_2009, guimera_missing_2009,
karrer_stochastic_2011,ball_efficient_2011,
reichardt_interplay_2011,decelle_asymptotic_2011}.  In addition to its
usefulness in this context, stochastic blockmodels serve as a general
framework which has many potential applications, such as the
parametrization of network topologies on which dynamical processes can
occur~\cite{peixoto_emergence_2012,peixoto_behavior_2012}, and in the
modelling of adaptive networks, where the topology itself can vary
according to dynamical rules~\cite{gross_adaptive_2008}.

The standard stochastic blockmodel
formulation~\cite{holland_stochastic_1983} assumes that all vertices
belonging to the same block are statistically indistinguishable, which
means that they all have the same expected degree. This restriction is
not very attractive for a general model, since many observed networks
show an extreme variation of degrees, even between vertices perceived to
be of the same block (or ``community''). Recently, this class of model
has been augmented by the introduction of the ``degree-corrected''
variant~\cite{karrer_stochastic_2011}, which incorporates such degree
variation, and was shown to be a much better model for many empirical
networks. With this modification, the stochastic blockmodel becomes more
appealing, since (except for the degrees) it only discards \emph{local
  scale} properties of the network topology (such as clustering, motifs,
etc.~\cite{newman_networks:_2010}), but can represent well arbitrary
\emph{global} or \emph{mesoscale} properties, such as
assortativity/dissortativity~\cite{newman_mixing_2003}, community
structure~\cite{girvan_community_2002, fortunato_community_2010},
bipartite and multipartite adjacency, and many others.

In this work, we focus on the microcanonical
entropy~\cite{bianconi_entropy_2008, anand_entropy_2009,
  bianconi_entropy_2009, anand_gibbs_2010} of stochastic blockmodel
ensembles, defined as $\mathcal{S}=\ln\Omega$, where $\Omega$ is the
number of graphs in the ensemble. This quantity has the traditional
interpretation of measuring the degree of ``order'' of a given ensemble,
which is more disordered (i.e. random) if the entropy is larger. It is
also a thermodynamic potential, which, in conjunction with other
appropriate quantities such as energy --- representing different sorts
of interactions, such as homophily in social
systems~\cite{castellano_statistical_2009} or robustness in biological
dynamical models~\cite{peixoto_emergence_2012} ---
 can be used to describe the equilibrium properties of evolved network
systems~\cite{strauss_general_1986, burda_statistical_2001,
park_statistical_2004, park_solution_2004, palla_statistical_2004,
burda_network_2004, biely_statistical_2006,
fronczak_fluctuation-dissipation_2006, fronczak_phase_2007,
jeong_construction_2007, foster_communities_2010,
peixoto_emergence_2012}.

From the entropy $\mathcal{S}$ one can directly derive the
log-likelihood function $\mathcal{L}=\ln\mathcal{P}$, where
$\mathcal{P}$ is the probability of observing a given network
realization, which is used often in the blockmodel literature. Assuming
that each graph in the ensemble is realized with the same probability,
$\mathcal{P}=1/\Omega$, we have simply that $\mathcal{L} =
-\mathcal{S}$. The log-likelihood can be used to infer the most likely
block structure which matches a given network data, and thus plays a
central role in the context of blockmodel detection. However, the
expressions for the log-likelihood $\mathcal{L}$, as they are often
derived in the stochastic blockmodel literature, do not allow one to
directly obtain the entropy, either because the they are expressed in
non-closed form~\cite{holland_stochastic_1983,fienberg_statistical_1985,
faust_blockmodels:_1992, anderson_building_1992,
nowicki_estimation_2001, reichardt_interplay_2011, ball_efficient_2011},
or because they only contain terms which depend on \emph{a posteriori}
partition of a sample network, with the remaining terms
neglected~\cite{bickel_nonparametric_2009, karrer_stochastic_2011,
ball_efficient_2011}.

In this work, we derive expressions for the entropy of elementary
variations of the blockmodel ensembles. The choice of microcanonical
ensembles permits the use of straightforward combinatorics, which
simplify the analysis. We consider both the traditional and
degree-corrected variants of the model, as well as their implementations
as ensembles of multigraphs (with parallel edges and self-loops allowed)
and simple graphs (no parallel edges or self-loops allowed). The
degree-corrected variants considered here represent a generalization of
the original definition~\cite{karrer_stochastic_2011}, since arbitrary
nearest-neighbours degree correlations are also allowed.  For the
degree-corrected variants, we consider the imposed degree sequence on
the vertices both as ``soft'' and ``hard'' constraints: When the degree
constraints are ``soft'', it is assumed that the imposed degree on each
vertex is only an average over the ensemble, and their values over
sampled realizations are allowed to fluctuate. With ``hard''
constraints, on the other hand, it is imposed that the degree sequence
is always the same on all samples of the ensemble. We also consider the
directed versions of all ensembles. These represent further refinements
of the original definition~\cite{karrer_stochastic_2011}, which
considered only undirected graphs with ``soft'' degree constraints.

The entropy expressions derived represent generalizations of several
expressions found in the literature for the case without block
structure~\cite{bender_asymptotic_1974, bender_asymptotic_1978,
wormald_models_1999, bianconi_entropy_2009}, which are easily recovered
by setting the number of blocks to one.

As a direct application of the derived entropy functions, we use them to
define a log-likelihood function $\mathcal{L}$, which can be used to
detect the most likely blockmodel partition which fits a given network
data. We show that these estimators work very well to detect block
structures in networks where there are intrinsic (as in the case of
simple graphs with broad degree distributions) or extrinsic degree
correlations. In particular, the expressions derived in this work
perform better for networks with broad degree distributions than the
sparse approximation derived in~\cite{karrer_stochastic_2011}, which may
result in suboptimal partitions.

This paper is divided as follows. In Sec.~\ref{sec:def} we define the
traditional and degree-corrected stochastic blockmodel ensembles.  In
Secs.~\ref{sec:simple} to~\ref{sec:hard} we systematically derive
analytical expressions for the most fundamental ensemble variants,
including simple graphs (Sec.~\ref{sec:simple}) and multigraphs
(Sec.~\ref{sec:multigraphs}), both the traditional and (soft)
degree-corrected versions, as well as the undirected and directed
cases. In Sec.~\ref{sec:hard} we obtain the entropy for the
degree-corrected ensembles with hard degree constraints, for the same
variants described in the other sections. In Sec.~\ref{sec:detection} we
apply the derived entropy expression for the soft degree-corrected
ensemble to the problem of blockmodel detection, by using it as a
log-likelihood function. [Readers more interested in the application to
blockmodel detection can read Secs.~\ref{sec:def}
to~\ref{sec:simple_soft}, and then move directly to
Sec.~\ref{sec:detection}.]  We finalize in Sec.~\ref{sec:conclusion}
with a conclusion.

\section{Traditional and degree-corrected blockmodels}\label{sec:def}

\begin{figure}
  \begin{minipage}{0.49\columnwidth}
    \begin{center}
      \includegraphics[width=\columnwidth]{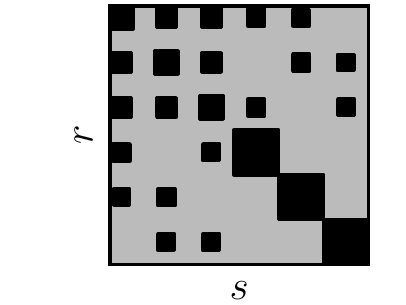}
    \end{center}
  \end{minipage}
  \begin{minipage}{0.49\columnwidth}
    \begin{center}
      \includegraphics[width=\columnwidth]{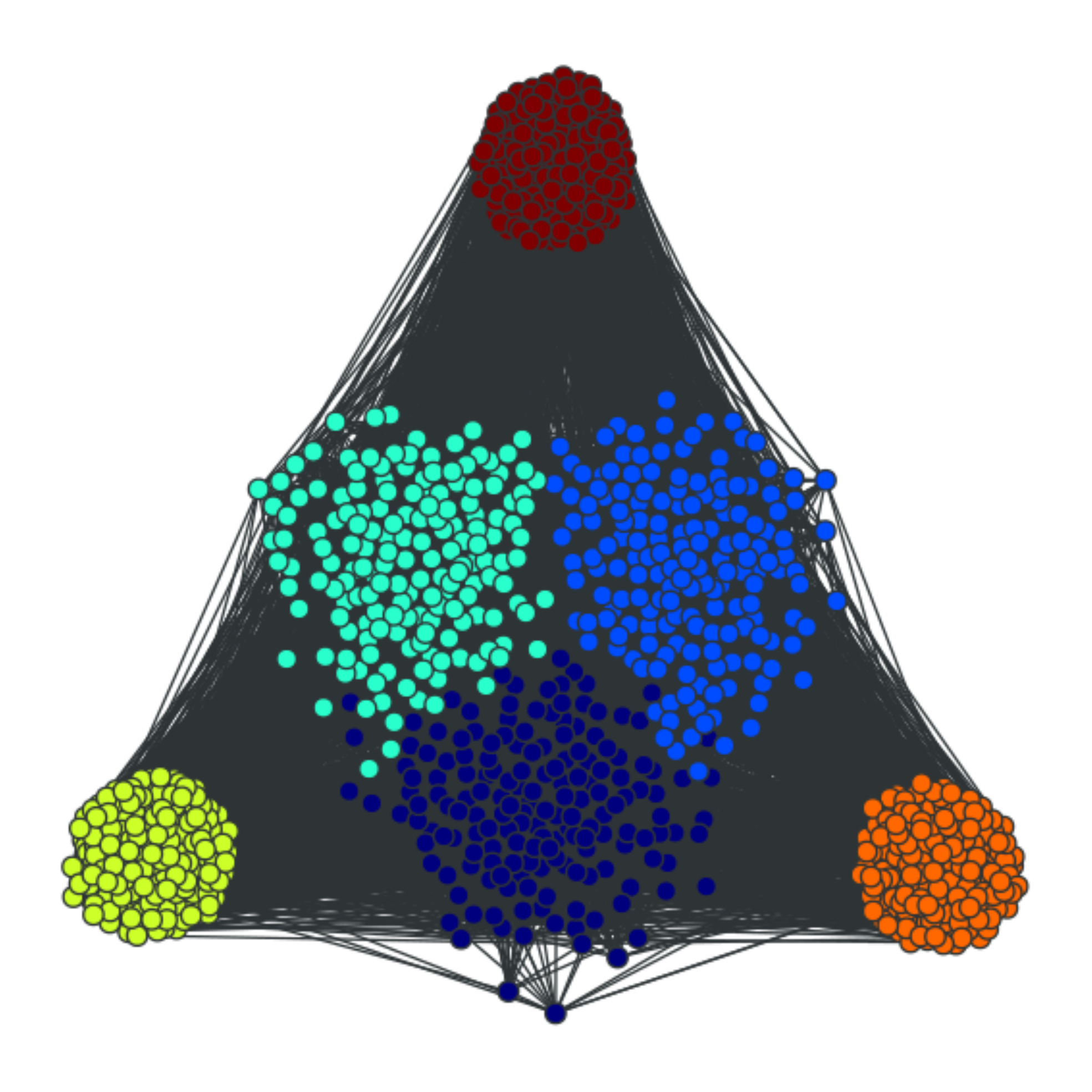}
    \end{center}
  \end{minipage}
  \caption{(Color online) Example of a traditional stochastic blockmodel with six
    blocks of equal size, and matrix $e_{rs}$ given on the left (each
    square is a matrix element, and its size corresponds to its
    magnitude). On the right is a sample of this ensemble with $10^3$
    vertices.\label{fig:example}}
\end{figure}

The traditional blockmodel ensemble is parametrized as follows: There
are $N$ vertices, partitioned into $B$ blocks, and $n_{r}$ is number of
vertices in block $r \in [0, B-1]$. The matrix $e_{rs}$ specifies the
number of edges between blocks $r$ and $s$, which are randomly
placed. As matter of convenience, the diagonal elements $e_{rr}$ are
defined as \emph{twice} the number of edges internal to the block $r$
(or equivalently, the number of ``half-edges''). An example of a
specific choice of parameters can be seen in Fig.~\ref{fig:example}.

This is a ``microcanonical'' formulation of the usual ``canonical'' form
which specifies instead the probability $w_{rs}$ of an edge occurring
between two vertices belonging to blocks $r$ and $s$, so that the
expected number of edges $e_{rs} = Ew_{rs}$ is allowed to fluctuate,
where $E$ is the total number of edges. If the nonzero values of
$e_{rs}$ are sufficiently large, these two ensembles become equivalent,
since in this case fluctuations around the mean value can be neglected.

The degree-corrected variant~\cite{karrer_stochastic_2011} further
imposes a degree sequence $\{k_i\}$ on each vertex $i \in [0, N-1]$ of
the network, which must be obeyed in addition to the block structure
specified by $n_r$ and $e_{rs}$. This restriction may be imposed in two
different ways. The first approach assumes these constraints are
``soft'', and each individual degree $k_i$ represents only the
\emph{average value} of the degree of vertex $i$ over all samples of the
ensemble~\cite{chung_average_2002,chung_connected_2002} (this is the
original ensemble defined in~\cite{karrer_stochastic_2011}). Here, we
will also consider a second approach which assumes the degree
constraints are ``hard'', and the imposed degree sequence must be
exactly the same in all samples of the ensemble. We will obtain the
entropy for both these ensembles in the following.

\section{Simple graph ensembles}\label{sec:simple}

\subsection{Standard stochastic blockmodel}

In simple graphs there can be at most only one edge between two
vertices. Therefore, we can enumerate the total number of different edge
choices between blocks $r$ and $s$ as,
\begin{equation}\label{eq:simle_edges}
  \Omega_{rs} = {n_rn_s \choose e_{rs}}, \qquad \Omega_{rr} = {{n_r \choose 2} \choose \frac{e_{rr}}{2}},
\end{equation}
which leads to the total number of graphs,
\begin{equation}
  \Omega = \prod_{r\ge s}\Omega_{rs}.
\end{equation}
The entropy is obtained by $\mathcal{S}_g=\ln\Omega$. Considering the
values of $n_r$ large enough so that Stirling's approximation can be
used, expressed as $\ln{N\choose m} \cong NH(m/N)$, where $H(x)$ is the
binary entropy function,
\begin{align} \label{eq:h}
  H(x) &= -x\ln x - (1-x)\ln(1-x)\\
       &= -x\ln x + x - \sum_{l=1}^{\infty} \frac{x^{l+1}}{l(l+1)},\label{eq:hexp}
\end{align}
we obtain the compact expression,
\begin{equation}\label{eq:ss}
  \mathcal{S}_g = \frac{1}{2} \sum_{rs} n_r n_s H\left(\frac{e_{rs}}{n_rn_s}\right).
\end{equation}
Eq.~\ref{eq:ss} has been derived by other means
in~\cite{bickel_nonparametric_2009} (expressed as a log-likelihood
function), for the canonical variant of the ensemble. Making use of the
series expansion given by Eq.~\ref{eq:hexp}, the entropy can be written
alternatively as
\begin{multline}\label{eq:ss_expanded}
  \mathcal{S}_g = E - \frac{1}{2} \sum_{rs} e_{rs} \ln\left(\frac{e_{rs}}{n_rn_s}\right) \\
  -\frac{1}{2} \sum_{rs} n_rn_s \sum_{l=1}^{\infty} \frac{1}{l(l+1)}\left(\frac{e_{rs}}{n_rn_s}\right)^{l+1},
\end{multline}
where $E=\sum_{rs}e_{rs}/2$ is the total number of edges in the network.
The terms in the last sum in the previous expression are of the order
$O(e^2_{rs}/n_rn_s)$. This number is typically of the order
$\sim\avg{k}^2$, where $\avg{k}$ is the average degree of the
network. Since the other terms of the expression are of order
$\sim\avg{k}{N}$, and one often has that $\avg{k} \ll N$, the last term
can be dropped, which leads to,
\begin{equation}\label{eq:sss}
  \mathcal{S}_{g} \cong E - \frac{1}{2} \sum_{rs} e_{rs} \ln\left(\frac{e_{rs}}{n_rn_s}\right).
\end{equation}
The last term of Eq.~\ref{eq:sss} is compatible with the equivalent
expression for the log-likelihood derived
in~\cite{karrer_stochastic_2011}. We note that while this limit can be
assumed in many practical scenarios, one can also easily imagine
ensembles which are ``globally sparse'' (i.e. $\avg{k} \ll N$), but
``locally dense'', with $\avg{k}_r = e_r/n_r \sim n_s $, for any two
blocks $r$, $s$ (with $e_r=\sum_se_{rs}$ being the total number of
half-edges adjacent to block $r$). In such scenarios Eq.~\ref{eq:sss}
will neglect potentially important contributions to the entropy, and
therefore Eqs.~\ref{eq:ss} or~\ref{eq:ss_expanded} should be used
instead.

As shown in~\cite{karrer_stochastic_2011}, the second term of
Eq.~\ref{eq:sss} can be slightly rewritten as the Kullback-Leibler
divergence~\cite{cover_elements_1991} between the actual and expected
distributions of block assignments at the opposing ends of randomly
chosen edges, where the expected distribution takes into account only
the size of each block. This can be interpreted as the amount of
additional information required to encode a given block partition, if
one assumes \emph{a priori} that the amount of edges incident to each
block is proportional to its size.

\subsubsection{Directed graphs}

The ensemble of directed blockmodels can be analysed in an analogous
fashion. The only differences is that for the directed version, the
matrix $e_{rs}$ can be asymmetric, and one needs to differentiate
between the number of edges leaving block $r$, $e^+_r = \sum_se_{rs}$,
and the number of edges arriving, $e^-_r = \sum_se_{sr}$. The number of
edge choices $\Omega_{rs}$ is given exactly as in
Eq.~\ref{eq:simle_edges}, the only difference being that one no longer
needs to differentiate the diagonal term, which in this case becomes
$\Omega_{rr} \equiv \Omega_{rs}|_{s=r}$. Since the matrix $e_{rs}$ is in
general asymmetric, the total number of graphs becomes the product over
all directed $r,s$ pairs,
\begin{equation}
  \Omega = \prod_{rs}\Omega_{rs}.
\end{equation}
Therefore the entropy becomes simply,
\begin{equation}\label{eq:dss}
  \mathcal{S}_g = \sum_{rs} n_r n_s H\left(\frac{e_{rs}}{n_rn_s}\right),
\end{equation}
which is identical to Eq.~\ref{eq:ss}, except for a factor $1/2$ (Note
that for directed graphs we define $e_{rr}$ as the number of edges
internal to block $r$, not twice this value as in the undirected
case). Naturally, the same alternative expression as in
Eq.~\ref{eq:ss_expanded} can be written, as well as the same
approximation as in Eq.~\ref{eq:sss}, which will be identical except for
a factor $1/2$.

\subsection{Degree-corrected ensembles with ``soft'' constraints}\label{sec:simple_soft}

Following~\cite{karrer_stochastic_2011}, we introduce degree variability
to the blockmodel ensemble defined previously, by imposing an
\emph{expected} degree sequence $\{\kappa_i\}$ on all vertices of the
graph, in addition to their block membership. Thus each individual
$\kappa_i$ represents only the \emph{average value} of the degree of
vertex $i$ over all samples of the ensemble. Such ``soft'' degree
constraints are relatively easy to implement, since one needs only to
extend the non degree-corrected version, simply by artificially
separating vertices with given imposed expected degrees into different
degree blocks. Thus, each existent block is labeled by a pair $(r,
\kappa)$, where the first value is the block label itself, and the
second is the expected degree label. In order for the label $(r,
\kappa)$ to be meaningful, we need to have intrinsically that
$e_{(r,\kappa)} =\sum_{s\kappa'}e_{(r,\kappa),(s,\kappa')} =\kappa
n_{(r,\kappa)}$, such that the average degree of vertices in block
$(r,\kappa)$ is exactly $\kappa$. This results in an ensemble with $KB$
blocks, where $K$ is the total number of different expected degrees,
$n_{(r,\kappa)}$ is the number of vertices in block $(r,\kappa)$, and
$e_{(r,\kappa),(s,\kappa')}$ is number of edges between $(r,\kappa)$ and
$(s,\kappa')$. Inserting this block structure into Eq.~\ref{eq:ss}, one
obtains
\begin{equation}\label{eq:ssd}
  \mathcal{S}_{gs} = \frac{1}{2} \sum_{\substack{r\kappa s\kappa'}} n_{(r,\kappa)} n_{(s,\kappa')} H\left(\frac{e_{(r,\kappa),(s,\kappa')}}{n_{(r,\kappa)} n_{(s,\kappa')}}\right).
\end{equation}
This ensemble accommodates not only blockmodels with arbitrary
(expected) degree sequences, but also with arbitrary degree
\emph{correlations}, since it is defined as a function of the full
matrix $e_{(r,\kappa),(s,\kappa')}$ (It is therefore a generalization of
the ensemble defined in~\cite{karrer_stochastic_2011}). However, it is
often more useful to consider the less-constrained ensemble where one
restricts only the total number of edges between blocks, irrespective of
their expected degrees,
\begin{equation}\label{eq:ec}
 e_{rs} = \sum_{\kappa\kappa'}e_{(r,\kappa),(s,\kappa')}.
\end{equation}
This can be obtained by maximizing the entropy $\mathcal{S}_{gs}$,
subject to this constraint. Carrying out this maximization, one arrives
at the following nonlinear system,
\begin{align}
  e_{(r,\kappa),(s,\kappa')} &= \frac{n_{(r,\kappa)} n_{(s,\kappa')}}{\exp(\lambda_{rs} + \mu_{r\kappa} + \mu_{s\kappa'}) + 1} \label{eq:fermi}\\
  e_{rs} &= \sum_{\kappa\kappa'}e_{(r,\kappa),(s,\kappa')} \label{eq:econst} \\
  \kappa n_{(r,\kappa)} &= \sum_{s\kappa'}e_{(r,\kappa),(s,\kappa')} \label{eq:kconst}
\end{align}
which must be solved for $\{e_{(r,\kappa),(s,\kappa')}, \lambda_{rs},
\mu_{r\kappa}\}$, where $\{\lambda_{rs}\}$ and $\{\mu_{r\kappa}\}$ are
Lagrange multipliers which impose the necessary constraints, described
by Eqs.~\ref{eq:econst} and \ref{eq:kconst},
respectively. Unfortunately, this system admits no general closed-form
solution. However, if one makes the assumption that $\exp(\lambda_{rs} +
\mu_{r\kappa} + \mu_{s\kappa'}) \gg 1$, one obtains the approximate
solution,
\begin{equation}\label{eq:classical_limit}
  e_{(r,\kappa),(s,\kappa')} \cong \frac{e_{rs}}{e_re_s} n_{(r,\kappa)} n_{(s,\kappa')} \kappa \kappa'.
\end{equation}
This is often called the ``sparse'' or ``classical''
limit~\cite{park_statistical_2004}, and corresponds to the limit where
intrinsic degree correlations between any two blocks $r$ and $s$ can be
neglected~\footnote{Because of the similarity of Eq.~\ref{eq:fermi} with
the Fermi-Dirac distribution in quantum mechanics, as well as the
analogy of the simple graph restriction with the Pauli exclusion
principle, this type of ensemble is sometimes called ``fermionic'', and
conversely the multigraph ensemble of Sec.~\ref{sec:multigraphs} is
called ``bosonic''~\cite{park_statistical_2004}. Note however that the
``classical'' limit represented by Eq.~\ref{eq:classical_limit} is still
insufficient to make these ensembles equivalent. This is only achieved
by the stronger sparsity condition given by Eq.~\ref{eq:sparse_limit}.}.
Eq.~\ref{eq:classical_limit} is intuitively what one expects for
uncorrelated degree-corrected blockmodels: The number of edges between
$(r,\kappa)$ and $(s,\kappa')$ is proportional to the number of edges
between the two blocks $e_{rs}$ and the degree values themselves,
$\kappa\kappa'$. Including this in Eq.~\ref{eq:ssd}, and using
Eq.~\ref{eq:hexp} one obtains,
\begin{multline}\label{eq:ssdu}
  \mathcal{S}_{gsu} \cong E - \sum_\kappa N_\kappa \kappa\ln \kappa - \frac{1}{2} \sum_{rs} e_{rs} \ln\left(\frac{e_{rs}}{e_re_s}\right) \\
  -\frac{1}{2} \sum_{rs} n_rn_s \sum_{l=1}^{\infty} \frac{1}{l(l+1)}\left(\frac{e_{rs}}{e_re_s}\right)^{l+1}\left<\kappa^{l+1}\right>_r\left<\kappa^{l+1}\right>_s,
\end{multline}
where $N_\kappa \equiv \sum_rn_{(r,\kappa)}$ is the total number of
vertices with expected degree $\kappa$, and $\avg{k^l}_r = \sum_{i\in
  r}\kappa_i^l / n_r$ is the $l$-th moment of the expected degree
sequence of vertices in block $r$. It is interesting to compare this
expression with the entropy $S_g$ for the non-degree corrected ensemble,
Eq.~\ref{eq:ss_expanded}. The importance of the terms in the last sum of
Eq.~\ref{eq:ssdu} will depend strongly on the properties of the expected
degree sequence $\{\kappa_i\}$. Irrespective of its average value, if
the higher moments $\left<\kappa^{l+1}\right>_r$ of a given block $r$
are large, so will be their contribution to the entropy. Therefore these
terms cannot be neglected \emph{a priori} for all expected degree
sequences, regardless of the values of the first moments
$\avg{k}_r$. Only if one makes the (relatively strong) assumption that,
\begin{equation}\label{eq:sparse_limit}
   n_rn_s \left(\frac{e_{rs}}{e_re_s}\right)^{l+1}\left<\kappa^{l+1}\right>_r\left<\kappa^{l+1}\right>_s \ll e_{rs},
\end{equation}
for any $l>0$, then Eq.~\ref{eq:ssdu} can be rewritten as,
\begin{equation}\label{eq:ssdu_sparse}
  \mathcal{S}_{gsu} \approx E - \sum_\kappa N_\kappa \kappa\ln \kappa - \frac{1}{2} \sum_{rs} e_{rs} \ln\left(\frac{e_{rs}}{e_re_s}\right).
\end{equation}
The last term of Eq.~\ref{eq:ssdu_sparse} is compatible with the
expression for the log-likelihood derived
in~\cite{karrer_stochastic_2011}, for the degree-corrected ensemble. It
is interesting to note that, in this limit, the block partition of the
network and the expected degree sequence contribute to independent terms
of the entropy. This means that the expected degrees can be distributed
in any way among the vertices of all blocks, without any entropic cost,
as long as the expected degree distribution is always the
same. Furthermore, as shown in~\cite{karrer_stochastic_2011}, the last
term of Eq.~\ref{eq:ssdu_sparse} can also be rewritten as the
Kullback-Leibler divergence between the actual and expected
distributions of block assignments at the opposing ends randomly chosen
edges, similarly to the non degree-corrected blockmodels. The main
difference now is that the expected distribution is expressed in terms
of the total number of half-edges $e_r$ leaving block $r$, instead of
the block size $n_r$. Equivalently, the last term corresponds (after
slight modifications) to the mutual information of block memberships at
the end of randomly chosen edges.

A typical situation where Eq.~\ref{eq:sparse_limit} holds is when the
expected degree sequence is such that the higher moments are related to
the first moment as $\left<\kappa^l\right>_r \sim
O(\left<\kappa\right>^l_r)$. This is the case, for instance, of expected
degrees distributed according to a Poisson. In this situation, the
left-hand side of Eq.~\ref{eq:sparse_limit} can be written as
$e_{rs}^{l+1}/(n_rn_s)^l$, and thus Eq.~\ref{eq:sparse_limit} holds when
$e_{rs}^2/n_rn_s \ll e_{rs}$, which is often the case for sparse graphs,
as discussed before for the non degree-corrected blockmodels. On the
other hand, if the expected degree distributions are broad enough, the
higher moments can be such that their contributions to the last term
cannot be neglected, even for sparse graphs. One particularly
problematic example are degree distributions which follow a power law,
$n_{(r,\kappa)} \propto \kappa^{-\gamma}$. Strictly speaking, for these
distributions all higher moments diverge, $\left<\kappa^l\right>_r \to
\infty$, for $l \ge \gamma - 1$. Of course, this divergence, in itself,
is inconsistent with the intrinsic constraints of simple graph
ensembles, since it would mean that there are expected degrees
$\kappa_i$ in the sequence which are larger than the network size, or
otherwise incompatible with the desired block structure. In order to
compute the moments correctly, one would need to consider more detailed
distributions, e.g. with structural cut-offs which depend on the network
size, or the sizes of the
blocks~\cite{boguna_cut-offs_2004}. Nevertheless, it is clear that in
such situations one would not be able to neglect the entropy terms
associated with the higher moments, since they can, in principle, be
arbitrarily large.

Note that certain choices of expected degree sequences are fundamentally
incompatible with Eq.~\ref{eq:classical_limit}, and will cause
Eq.~\ref{eq:ssdu} to diverge. If one inserts
Eq.~\ref{eq:classical_limit} into Eq.~\ref{eq:ssd}, the term inside the
sum becomes $H\left(e_{rs}\kappa\kappa'/e_re_s\right)$. Since the binary
entropy function $H(x)$ is only defined for arguments in the range $0\le
x \le 1$, then Eq.~\ref{eq:ssdu_sparse} will only converge if the
following holds,
\begin{equation}\label{eq:kk_const}
  \kappa\kappa' \le \frac{e_re_s}{e_{rs}},
\end{equation}
for all $\kappa$, $\kappa'$ belonging to blocks $r$ and $s$,
respectively. If Eq.~\ref{eq:kk_const} is not fulfilled, then
Eq.~\ref{eq:classical_limit} cannot be used as an approximation for the
solution of the system in Eqs.~\ref{eq:fermi} to~\ref{eq:kconst}, and
consequently Eq.~\ref{eq:ssdu} becomes invalid. Note that even if
Eq.~\ref{eq:kk_const} is strictly fulfilled, it may also be the case
that Eq.~\ref{eq:classical_limit} is a bad approximation, which means
there will be strong intrinsic inter-block dissortative degree
correlations~\cite{park_origin_2003, johnson_entropic_2010}. A
sufficient condition for the applicability of Eq.~\ref{eq:ssdu} would
therefore be $\kappa\kappa' \ll e_re_s/e_{rs}$, for all $\kappa$,
$\kappa'$ belonging to blocks $r$ and $s$, respectively. However, it is
important to emphasize that even if Eq.~\ref{eq:classical_limit} is
assumed to be a good approximation, it only means that the intrinsic
degree correlations between any given block pair $r,s$ can be neglected,
but the entropic cost of connecting to a block with a broad degree
distribution is still reflected in the last term of
Eq.~\ref{eq:ssdu}. This captures one important entropic effect of broad
distributions, which can be important, e.g. in inferring block
structures from empirical data, as will be shown in
Sec.~\ref{sec:detection}.

\subsubsection{Directed graphs}

The directed degree-corrected variants can be analysed in analogous
fashion, by separating vertices into blocks depending on their expected
in- and out-degrees, leading to block labels given by
$(r,\kappa^-,\kappa^+)$, which are included directly into
Eq.~\ref{eq:dss} above, which leads to an expression equivalent to
Eq.~\ref{eq:ssd}, which is omitted here for brevity. The ``classical''
limit can also be taken, which results in the expression,
\begin{equation}\label{eq:dcorr}
  e_{(r,\kappa^-,\kappa^+),(s,{\kappa'}^-,{\kappa'}^+)} \cong \frac{e_{rs}}{e^+_re^-_s} n_{(r,\kappa^-,\kappa^+)} n_{(s,{\kappa'}^-,{\kappa'}^+)} \kappa^+ {\kappa'}^-,
\end{equation}
which if inserted into the degree-corrected entropy expression leads
to,
\begin{multline}\label{eq:dsmdu}
  \mathcal{S}_{gsu} \cong E - \sum_{\kappa^+}N_{\kappa^+} \kappa^+\ln \kappa^+ - \sum_{\kappa^-}N_{\kappa^-} \kappa^-\ln \kappa^-\\
  - \sum_{rs} e_{rs} \ln\left(\frac{e_{rs}}{e_r^+e_s^-}\right) \\
  - \sum_{rs} n_rn_s \sum_{l=1}^{\infty} \frac{1}{l(l+1)}\left(\frac{e_{rs}}{e_r^+e_s^-}\right)^{l+1}\left<(\kappa^+)^{l+1}\right>_r\left<(\kappa^-)^{l+1}\right>_s.
\end{multline}
The same caveats as in the undirected case regarding the suitability of
Eq.~\ref{eq:dcorr}, and consequently the validity of Eq.~\ref{eq:dsmdu},
apply.

\section{Multigraph ensembles}\label{sec:multigraphs}

We now consider the situation where multiple edges between the same
vertex pair are allowed.  The total number of different edge choices
between blocks $r$ and $s$ now becomes,
\begin{equation}\label{eq:multi_edges}
  \Omega_{rs} = \multiset{n_rn_s}{e_{rs}}, \qquad \Omega_{rr} = \left(\!\!{\left(\!\!{n_r \choose 2}\!\!\right) \choose \frac{e_{rr}}{2}}\!\!\right),
\end{equation}
where $\multiset{N}{m} = {N + m - 1 \choose m}$ is the total number of
$m$-combinations \emph{with repetition} from a set of size $N$.  Like
for simple graphs, total number of graphs is given by the total number
of vertex pairings between all blocks,
\begin{equation}
  \Omega = \prod_{r\ge s}\Omega_{rs},
\end{equation}
which leads to the entropy,
\begin{equation}\label{eq:sm_exact}
  \mathcal{S}_{m} = \frac{1}{2} \sum_{rs} (n_rn_s + e_{rs}) H\left(\frac{n_rn_s}{n_rn_s + e_{rs}}\right),
\end{equation}
where $H(x)$ is the binary entropy function (Eq.~\ref{eq:h}), as before.
If we consider the more usual case when $e_{rs} \le n_rn_s$, we can
expand this expression as,
\begin{multline}\label{eq:sm_expanded}
  \mathcal{S}_{m} = E - \frac{1}{2} \sum_{rs} e_{rs} \ln\left(\frac{e_{rs}}{n_rn_s}\right) \\
  + \sum_{rs}n_rn_s\sum_{l=1}^\infty\frac{(-1)^{l+1}}{l(l+1)}\left(\frac{e_{rs}}{n_rn_s}\right)^{l+1}.
\end{multline}
This is very similar to Eq.~\ref{eq:ss_expanded} for the simple graph
ensemble, with the only difference being the alternating sign in the
last term. In the sparse limit, the last term can also be dropped, which
leads to,
\begin{equation}\label{eq:sm}
  \mathcal{S}_{m} \cong E - \frac{1}{2} \sum_{rs} e_{rs} \ln\left(\frac{e_{rs}}{n_rn_s}\right).
\end{equation}
In this limit, the entropy is identical to the simple graph ensemble,
since the probability of observing multiple edges vanishes.

\subsubsection{Directed graphs}

Like for the simple graph case, the entropy for directed multigraphs can
be obtained with only small modifications. The number of edge choices
$\Omega_{rs}$ is given exactly as in Eq.~\ref{eq:multi_edges}, the only
difference being that one no longer needs to differentiate the diagonal
term, which in this case becomes $\Omega_{rr} \equiv
\Omega_{rs}|_{s=r}$. Since the matrix $e_{rs}$ is in general asymmetric,
the total number of graphs becomes the product over all directed $r,s$
pairs,
\begin{equation}
  \Omega = \prod_{rs}\Omega_{rs}.
\end{equation}
Therefore the entropy becomes simply,
\begin{equation}\label{eq:dms}
  \mathcal{S}_g = \sum_{rs} (n_rn_s + e_{rs}) H\left(\frac{n_rn_s}{n_rn_s + e_{rs}}\right)
\end{equation}
which is identical to Eq.~\ref{eq:sm_exact}, except for a factor $1/2$
(Note that for directed graphs we define $e_{rr}$ as the number of edges
internal to block $r$, not twice this value as in the undirected
case). Again, the same alternative expression as in
Eq.~\ref{eq:sm_expanded} can be written, as well as the same
approximation as in Eq.~\ref{eq:sm}, which will be identical except for
a factor $1/2$.

\subsection{Degree-corrected ensembles with ``soft''  constraints}

We proceed again analogously to the simple graph case, and impose that
each block is labeled by a pair $(r, \kappa)$, where the first value is
the block label itself, and the second is expected the degree
block. Using this labeling we can write the full entropy from
Eq.~\ref{eq:sm_exact} as,
\begin{multline}\label{eq:smd}
  \mathcal{S}_{ms} = \frac{1}{2} \sum_{\substack{r\kappa s\kappa'}} (n_{(n,\kappa)}n_{(s,\kappa')} + e_{(r,\kappa),(s,\kappa')}) \times \\
  H\left(\frac{n_{(r,\kappa)} n_{(s,\kappa')}}{n_{(r,\kappa)} n_{(s,\kappa')} + e_{(r,\kappa),(s,\kappa')}}\right).
\end{multline}
Like for the simple graph case, this is a general ensemble which allows
for arbitrary degree correlations. The ``uncorrelated'' ensemble
is obtained by imposing the constraint given by Eq.~\ref{eq:ec}, and
maximizing $\mathcal{S}_{ms}$, which leads to the following nonlinear
system,
\begin{align}
  e_{(r,\kappa),(s,\kappa')} &= \frac{n_{(r,\kappa)} n_{(s,\kappa')}}{\exp(\lambda_{rs} + \mu_{r\kappa} + \mu_{s\kappa'}) - 1} \\
  e_{rs} &= \sum_{\kappa\kappa'}e_{(r,\kappa),(s,\kappa')} \\
  \kappa n_{(r,\kappa)} &= \sum_{s\kappa'}e_{(r,\kappa),(s,\kappa')}
\end{align}
which must be solved for $\{e_{(r,\kappa),(s,\kappa')}, \lambda_{rs},
\mu_{r\kappa}\}$. where $\{\lambda_{rs}\}$ and $\{\mu_{r\kappa}\}$ are
Lagrange multipliers which impose the necessary constraints. Like for
the simple graph case, this system does not have a closed form solution,
but one can consider the same ``classical'' limit, $\exp(\lambda_{rs} +
\mu_{r\kappa} + \mu_{s\kappa'}) \gg 1$, which leads to
Eq.~\ref{eq:classical_limit}. Inserting it in Eq.~\ref{eq:smd}, and
using the series expansion given by Eq.~\ref{eq:hexp}, the entropy can
be written as,
\begin{multline}
  \mathcal{S}_{msu} \cong E - \sum_\kappa N_\kappa \kappa\ln \kappa - \frac{1}{2} \sum_{rs} e_{rs} \ln\left(\frac{e_{rs}}{e_re_s}\right) \\
  +\frac{1}{2} \sum_{rs} n_rn_s \sum_{l=1}^{\infty} \frac{(-1)^{l+1}}{l(l+1)}\left(\frac{e_{rs}}{e_re_s}\right)^{l+1}\left<\kappa^{l+1}\right>_r\left<\kappa^{l+1}\right>_s.
\end{multline}
Again, the difference from the simple graph ensemble is only the
alternating sign in the last term. If one takes the sparse limit, the
above equation is approximated by Eq.~\ref{eq:ssdu_sparse}, since in
this case both ensembles become equivalent.

\subsubsection{Directed graphs}

Directed multigraphs can be analysed in the same way, by using block
labels given by $(r,\kappa^-,\kappa^+)$, which are included into
Eq.~\ref{eq:dms} above, which leads to an expression equivalent to
Eq.~\ref{eq:smd}, which is omitted here for brevity. The ``classical''
limit can also be taken, which results in Eq.~\ref{eq:dcorr}, as for
simple graphs. Inserting it into the degree-corrected entropy expression
leads finally to,
\begin{multline}\label{eq:dmdu}
  \mathcal{S}_{msu} \cong E - \sum_{\kappa^+}N_{\kappa^+} \kappa^+\ln \kappa^+ - \sum_{\kappa^-}N_{\kappa^-} \kappa^-\ln \kappa^-\\
  - \sum_{rs} e_{rs} \ln\left(\frac{e_{rs}}{e_r^+e_s^-}\right) \\
  + \sum_{rs} n_rn_s \sum_{l=1}^{\infty} \frac{(-1)^{l+1}}{l(l+1)}\left(\frac{e_{rs}}{e_r^+e_s^-}\right)^{l+1}\left<(\kappa^+)^{l+1}\right>_r\left<(\kappa^-)^{l+1}\right>_s,
\end{multline}
which is once again similar to the simple graph ensemble, except for the
alternating sign in the last term.  The same caveats as in the simple
graph case regarding the suitability of Eq.~\ref{eq:dcorr}, and
consequently the validity of Eq.~\ref{eq:dmdu}, apply.

\section{Degree-corrected ensembles with ``hard'' constraints}\label{sec:hard}

For the case of ``hard'' degree constraints we cannot easily adapt any
of the counting schemes used so far. In fact, for the simpler case of a
single block ($B=1$), which is the ensemble of random graphs with a
prescribed degree sequence~\cite{bender_asymptotic_1974,
  bender_asymptotic_1978, mckay_asymptotic_1990, wormald_models_1999, bianconi_entropy_2009},
there is no known asymptotic expression for the entropy which is
universally valid. Even the simpler asymptotic counting of graphs with
an uniform degree sequence ($k_i=k$ for all $i$) is an open problem in
combinatorics~\cite{wormald_models_1999}. All known expressions are
obtained by imposing restrictions on the largest degree of the
sequence~\cite{bender_asymptotic_1974, bender_asymptotic_1978,
  mckay_asymptotic_1990, wormald_models_1999}, such that $k_i \ll N$,
where $N$ is the number of vertices in the graph~\footnote{Note that not
all degree sequences are allowed in the first place, since they must be
\emph{graphical}~\cite{erdos_graphs_1960,del_genio_all_2011}. Imposing a
block structure further complicates things, since the graphical
condition needs to be generalized to blockmodels. We will not pursue
this here, as we consider only the sufficiently sparse situation, where
this issue can be neglected.}. Here we make similar assumptions, and
obtain expressions which are valid only for such sparse limits, in
contrast to the other expressions calculated so far. The approach we
will take is to start with the ensemble of
\emph{configurations}~\cite{bollobas_probabilistic_1980}, which contains
all possible half-edge pairings obeying a degree sequence. Each
configuration (i.e. a specific pairing of half-edges) corresponds to
either a simple graph or a multigraph, but any given simple graph or
multigraph will correspond to more than one configuration. Knowing the
total number of configurations $\Omega^c_{rs}$ between blocks $r$ and
$s$, the total number $\Omega_{rs}$ of edge choices corresponding to
distinct graphs can then be written as,
\begin{equation}
  \Omega_{rs} = \Omega^c_{rs}\Xi_{rs},
\end{equation}
where $\Xi_{rs}$ is the fraction of configurations which correspond to
distinct simple graphs or multigraphs.

Although counting configurations and graphs are different, and so will
be the corresponding entropies, there are some stochastic processes and
algorithms which generate fully random configurations, instead of
graphs. Perhaps the most well known example is the configurational
model~\cite{newman_random_2001, newman_structure_2003}, which is the
ensemble of all configurations which obey a prescribed degree
sequence. A sample from this ensemble can be obtained with a simple
algorithm which randomly matches
half-edges~\cite{newman_random_2001}. If one rejects multigraphs which
are generated by this algorithm, one has a (possibly very inneficient)
method of generating random graphs with a prescribed degree sequence,
since each simple graph will be generated by the same number of
configurations, which is given by $\prod_i k_i!$. However, the same is
not true if one attempts to generate multigraphs, since they will not be
equiprobable~\cite{king_comment_2004}, as will be discussed in
Sec.~\ref{sec:hard_multigraphs} below.

A central aspect of computing $\Xi_{rs}$ is the evaluation of the
probability of obtaining multiple edges. If we isolate a given pair
$i,j$ of vertices, which belong to block $r$ and $s$, respectively, we
can write the probability of there being $m$ parallel edges between them
as,
\begin{equation}
  P^{rs}_{ij}(m) = \frac{{k'_j \choose m} {e_{rs} - k'_j \choose k'_i - m}}{{e_{rs} \choose k'_i}}
\end{equation}
which is the hypergeometric distribution, since each half-edge can only
be paired once (i.e. there can be no replacement of half-edges). In the
above expression, the degrees $k'_i$ and $k'_j$ reflect the number of
edges in each vertex which lie between blocks $r$ and $s$, which can be
smaller than the total degrees, $k_i$ and $k_j$.  In general, this
expression \emph{is not valid independently for all pairs} $i,j$, since
the pairing of two half-edges automatically restricts the options
available for other half-edges belonging to different vertex
pairs. However, in the limit where the largest degrees in each block are
much smaller than the total number of vertices in the same blocks, we
can neglect such interaction between different placements, since the
number of available options is always approximately the same. This is
not a rigorous assumption, but it is known to produce results which are
compatible with more rigorous (and laborious)
analysis~\cite{bender_asymptotic_1978, bianconi_entropy_2009}. In the
following we compute the number of configurations and the approximation
of $\Xi_{rs}$ for simple graphs and multigraphs, using this assumption.

\subsection{Configurations}

For a given block $r$, the number of different half-edge pairings which
obey the desired block structure determined by $e_{rs}$ is given by,
\begin{equation}
\Omega_r = \frac{e_r!}{\prod_s e_{rs}!}.
\end{equation}
The above counting only considers to which block a given half-edge is
connected, not specific half-edges. The exact number of different
pairings between two blocks is then given simply by,
\begin{equation}
  \Omega_{rs} = e_{rs}!, \qquad \Omega_{rr} = (e_{rr} - 1)!!.
\end{equation}
Note that the above counting differentiates between permutations of the
out-neighbours of the same vertex, which are all equivalent
(i.e. correspond to the same graph). This can be corrected in the full
number of pairings,
\begin{equation}\label{eq:confs}
  \Omega = \frac{\prod_r\Omega_r\prod_{s\ge r}\Omega_{rs}}{\prod_k (k!)^{N_k}},
\end{equation}
where the denominator discounts all equivalent permutations of
out-neighbours. Note that the above counting still does not account for
the total number of simple graphs, since multiedges are still
possible. Multigraphs are also not counted correctly, since for each
occurrence of $m$ multiedges between a given vertex pair, the number of
different edge pairings which are equivalent decreases by a factor
$m!$~\cite{king_comment_2004, newman_networks:_2010}. These corrections
are going to be considered in the next sections.  Taking the logarithm
of Eq.~\ref{eq:confs}, and using Stirling's approximation, one obtains,
\begin{equation}\label{eq:smdh}
  \mathcal{S}_c = -E - \sum_kN_k\ln k! - \frac{1}{2} \sum_{rs} e_{rs} \ln\left(\frac{e_{rs}}{e_re_s}\right).
\end{equation}
It is interesting to compare this expression with the one obtained for
soft degree-constraints in the sparse limit,
Eq.~\ref{eq:ssdu_sparse}. The entropy difference between the two
ensembles depends only on the degree sequence,
\begin{equation}
  \mathcal{S}_{gsu} - \mathcal{S}_{c} = 2E + \sum_kN_k\ln k! - \sum_\kappa N_\kappa \kappa \ln \kappa.
\end{equation}
This difference disappears if the individual degrees are large enough so
that Stirling's approximation can be used, i.e. $\ln k! \approx k\ln k -
k$, and we have that $k_i = \kappa_i$ for all vertices. Thus, in the
sparse limit, but with sufficiently large degrees, the simple graph and
multigraph ensembles with soft constraints, and the configuration
ensemble with hard constraints become equivalent~\footnote{The
  difference between ensembles with ``hard'' and ``soft'' degree
  constraints is analyzed in detail in~\cite{anand_gibbs_2010} for the
  case without block structures.}.

\subsubsection{Directed configurations}
When counting directed configurations, we no longer need to discriminate
the diagonal terms of the $\Omega_{rs}$ matrix, which become
$\Omega_{rr} \equiv e_{rr}!$. Since the matrix $e_{rs}$ is in general
asymmetric, the total number of configurations becomes,
\begin{equation}
  \Omega = \frac{\prod_r\Omega_r\prod_{rs}\Omega_{rs}}{\prod_{k^+} (k^+!)^{N_{k^+}}\prod_{k^-} (k^-!)^{N_{k^-}}},
\end{equation}
which includes the correction for the permutations of in- and
out-degrees. This leads to the entropy,
\begin{multline}
  \mathcal{S}_{cd} = -E - \sum_{k^+}N_{k^+}\ln {k^+}! - \sum_{k^-}N_{k^-}\ln {k^-}! \\ -\sum_{rs} e_{rs} \ln\left(\frac{e_{rs}}{e_r^+e_s^-}\right).
\end{multline}

\subsection{Simple graphs}

Following~\cite{bianconi_entropy_2009}, if we proceed with the
assumption outlined above that $P^{rs}_{ij}(m)$ are independent
probabilities of there being $m$ edges between vertices $i$ and $j$, we
can write the probability $\Xi_{rs}$ that a configuration corresponds to
a simple graph as,
\begin{align}
  \Xi_{rs} &\approx \prod_{ij}[P^{rs}_{ij}(0) + P^{rs}_{ij}(1)] \\
  \Xi_{rr} &\approx \prod_{i>j} [P^{rr}_{ij}(0) + P^{rr}_{ij}(1)] \times \prod_iP^r_{nl}(i),
\end{align}
where the product is taken over all vertex pairs $i,j$, belonging to
blocks $r$ and $s$, respectively, and $P^r_{nl}(i)$ is the probability
of there being no self-loops attached to vertex $i$, belonging to block
$r$. This is given by computing the probability that all $k_i$ half-edge
placements are not self-loops,
\begin{align}
  P^r_{nl}(i) &= \frac{e_{rr} - k'_i}{e_{rr}-1} \frac{e_{rr} - k'_i-1}{e_{rr}-3} \cdots \frac{e_{rr} - 2k'_i+1}{e_{rr} - 2k'_i+1} \\
           &= \frac{(e_{rr}-k'_i)!(e_{rr}-2k'_i-1)!!}{(e_{rr}-2k'_i)!(e_{rr}-1)!!}, \label{eq:pnl}
\end{align}
where we also make the assumption that these probabilities are
independent for all vertices.  We proceed by applying Stirling's
approximation up to logarithmic terms,
i.e.  $\ln{x!} \approx (x-1/2)\ln x - x$ , and expanding the
probabilities in powers of $1/e_{rs}$, leading to,
\begin{multline}
  \ln[P^{rs}_{ij}(0) + P^{rs}_{ij}(1)] \approx -\frac{2}{e_{rs}^2} {k'_i \choose 2} {k'_j \choose 2} + O(1/e_{rs}^3)
\end{multline}
and,
\begin{equation}
  \ln P^r_{nl}(i) \approx - \frac{1}{e_{rs}}{k'_i \choose 2} + O(1/e_{rr}^2).
\end{equation}
As mentioned before, the degrees $k'_i$ and $k'_j$ in the expression
above are number of edges in each vertex which lie between blocks $r$
and $s$. Since the total degrees $k_i$ and $k_j$ are assumed to be much
smaller than the number of half-edges leaving each block, we can
consider $k'_i$, for $i\in r$, to be a binomially distributed random
number in the range $[0,k_i]$, with a probability $e_{rs}/e_r$. We can
therefore write $\avg{k'_i} = k_i e_{rs} / e_r$, and $\avg{{k'_i}^2} =
k_i(k_i-1) e_{rs}^2 / e_r^2$, where the average is taken over all
vertices with the same degree and in the same block $r$. Putting it all
together we obtain an expression for the entropy which reads,
\begin{multline}\label{eq:shdu}
  \mathcal{S}_{ghu} \approx -E - \sum_kN_k\ln k! - \frac{1}{2} \sum_{rs} e_{rs} \ln\left(\frac{e_{rs}}{e_re_s}\right) \\
  - \frac{1}{4} \sum_{rs}\frac{n_rn_se_{rs}^2}{e_r^2e_s^2}\left(\avg{k^2}_r-\avg{k}_r\right)\left(\avg{k^2}_s-\avg{k}_s\right) \\
  - \frac{1}{2} \sum_r \frac{n_re_{rr}}{e^2_r}\left(\avg{k^2}_r-\avg{k}_r\right),
\end{multline}
where $\avg{k}_r = \sum_{i \in r}k_i / n_r$ and $\avg{k^2}_r = \sum_{i
  \in r}k^2_i / n_r$.

If we make $B=1$, the ensemble is equivalent to fully random graphs with
an imposed degree sequence. In this case, Eq.~\ref{eq:shdu} becomes
identical to the known expression derived
in~\cite{bender_asymptotic_1978}, for the limit $k_i \ll N$ (which is
known to be valid for $\max(\{k_i\}) \sim
o(\sqrt{N})$~\cite{janson_probability_2009}). This expression is also
compatible with the one later derived in~\cite{bianconi_entropy_2009}
(except for a trivial constant). Therefore we have obtained an
expression which is fully consistent with the known special case without
block structure.

It is interesting to compare Eq.~\ref{eq:shdu} with the equivalent
expression for the case with soft degree constraints,
Eq.~\ref{eq:ssdu}. Eq.~\ref{eq:shdu} is less complete than
Eq.~\ref{eq:ssdu} since it contains terms of order comparable only to
the first term of the sum in Eq.~\ref{eq:ssdu}. Furthermore, in
Eq.~\ref{eq:shdu} the last terms involve the difference
$\avg{k^2}_r-\avg{k}_r$, instead of the second moment $\avg{k^2}_r$, as
in Eq.~\ref{eq:ssdu}. [It is worth noting that Eq.~\ref{eq:shdu} passes
the ``sanity check'' of making $\avg{k^2}_r=\avg{k}_r$, which is only
possible for the uniform degree sequence $k_i = 1$, in which case no
parallel edges are possible, and the entropy becomes identical to the
ensemble of configurations, Eq.~\ref{eq:smdh}.] Thus we can conclude
that the two ensembles (with soft and hard constraints) are only
equivalent in the sufficiently sparse case when the differences in the
remaining higher order terms in Eq.~\ref{eq:ssdu} can be neglected, and
when the degrees are large enough (or the distributions broad enough) so
that $\avg{k^2}_r \gg \avg{k}_r$, and the self-loop term can also be
discarded.

\subsubsection{Directed graphs}

For directed graphs one can proceed stepwise with an analogous
calculation, with the only difference that the probability of self-loops
in this case involves the in- and out-degree of the same vertex, and can
be obtained by the hypergeometric distribution,
\begin{align}
  P^r_{nl}(i) &= {e_{rr} - k^+_i \choose k^-_i}\left/{e_{rr} \choose k_i^-}\right. \\
             &\approx \exp\left(-\frac{k_i^+k_i^-}{e_{rr}} + O(1/e_{rr}^2)\right).
\end{align}
The analogous expression to Eq.~\ref{eq:shdu} then becomes,
\begin{multline}
  \mathcal{S}_{ghu} \approx \mathcal{S}_{cd}
  - \frac{1}{2} \sum_{rs}\frac{n_rn_se_{rs}^2}{(e_r^+)^2(e_s^-)^2}\left(\avg{(k^+)^2}_r-\avg{k^+}_r\right)\times\\
  \left(\avg{(k^-)^2}_s-\avg{k^-}_s\right)
  - \sum_r \frac{n_re_{rr}}{e^+_re^-_r}\avg{k^+k^-}_r.
\end{multline}
Similarly to Eq.~\ref{eq:shdu}, if we make $B=1$, we recover the known
expression derived in~\cite{bender_asymptotic_1978} for the number of
directed simple graphs with imposed in/out-degree sequence, obtained for
the limit $k^{-/+} \ll N$.

\subsection{Multigraphs}\label{sec:hard_multigraphs}

All configurations which are counted in Eq.~\ref{eq:smdh} are
multigraphs, but not all multigraphs are counted the same number of
times. More precisely, for each vertex pair of a given graph with $m$
edges between them, the number of configurations which generate this
graph is smaller by a factor of $m!$, compared to a simple graph of the
same ensemble~\cite{king_comment_2004, newman_networks:_2010}. This
means that the denominator of Eq.~\ref{eq:confs} overcounts the number
of equivalent configurations for graphs with multiedges. Hence,
similarly to the simple graph case, we can calculate the correction
$\Xi_{rs}$ as,
\begin{align}
  \Xi_{rs} &\approx \prod_{ij}\avg{m!}^{rs}_{ij},\\
  \Xi_{rr} &\approx \prod_{i>j}\avg{m!}^{rr}_{ij} \times  \prod_{i}\avg{(2m)!!}^{r}_{i},
\end{align}
where $\avg{m!}^{rs}_{ij} = \sum_{m=0}^{\infty} m!P^{rs}_{ij}(m)$ is the
average correction factor, and $\avg{(2m)!!}^{r}_{i}=\sum_{m=0}^{\infty}
(2m)!!\hat{P}^{r}_{i}(m)$ accounts for the parallel self-loops, with
$\hat{P}^{r}_{i}(m)$ being the probability of observing $m$ parallel
self-loops on vertex $i$, belonging to block $r$. It is easy to see that
$\hat{P}^{r}_{i}(m=0)=P^r_{nl}$, given by Eq.~\ref{eq:pnl},
$\hat{P}^{r}_{i}(m=1)\cong {k_i \choose 2}/e_{rr} + O(1/e_{rr}^2)$ and
$\hat{P}^{r}_{i}(m>1)\sim O(1/e_{rr}^m)$. We proceed by applying
Stirling's approximation up to logarithmic terms, i.e.  $\ln{x!} \approx
(x-1/2)\ln x - x$, and expanding the sum in powers of $1/e_{rs}$, which
leads to,
\begin{align}
  \ln\avg{m!}^{rs}_{ij} &\approx \frac{2}{e_{rs}^2} {k'_i \choose 2} {k'_j \choose 2} + O(1/e_{rs}^3),\\
  \ln\avg{(2m)!!}^{r}_{i} &\approx \frac{1}{e_{rr}} {k'_i \choose 2} + O(1/e_{rr}^2).
\end{align}
Using that $\avg{k'_i} = k_i e_{rs} / e_r$, and $\avg{{k'_i}^2} =
k_i(k_i-1) e_{rs}^2 / e_r^2$, and putting it all together we obtain an
expression for the entropy which reads,
\begin{multline}\label{eq:mhdu}
  \mathcal{S}_{mhu} \approx -E - \sum_kN_k\ln k! - \frac{1}{2} \sum_{rs} e_{rs} \ln\left(\frac{e_{rs}}{e_re_s}\right) \\
  + \frac{1}{4} \sum_{rs}\frac{n_rn_se_{rs}^2}{e_r^2e_s^2}\left(\avg{k^2}_r-\avg{k}_r\right)\left(\avg{k^2}_s-\avg{k}_s\right)\\
  + \frac{1}{2} \sum_r \frac{n_re_{rr}}{e^2_r}\left(\avg{k^2}_r-\avg{k}_r\right),
\end{multline}
where $\avg{k}_r = \sum_{i \in r}k_i / n_r$ and $\avg{k^2}_r = \sum_{i
\in r}k^2_i / n_r$. This expression is very similar to the one obtained
for the simple graph ensemble, except for the sign of the last two
terms.

Again if we make $B=1$, the ensemble is equivalent to fully random
multigraphs with an imposed degree sequence. In this case,
Eq.~\ref{eq:mhdu} becomes identical to the known expression derived
in~\cite{wormald_asymptotic_1981}, for the limit $k_i \ll N$. It also
corresponds to the expression derived in~\cite{bender_asymptotic_1978},
which does not include the last term, since in that work parallel
self-edges are effectively counted as contributing degree one to a
vertex, instead of two as is more typical.

\subsubsection{Directed multigraphs}

For directed graphs one can proceed stepwise with an analogous
calculation, which leads to,
\begin{multline}
  \mathcal{S}_{ghu} \approx \mathcal{S}_{cd}
  + \frac{1}{2} \sum_{rs}\frac{n_rn_se_{rs}^2}{(e_r^+)^2(e_s^-)^2}\times \\
  \left(\avg{(k^+)^2}_r-\avg{k^+}_r\right)\left(\avg{(k^-)^2}_s-\avg{k^-}_s\right).
\end{multline}
Note that in this case the calculation of the correction term for
self-loops is no different than other parallel edges, and hence there is
no self-loop term as in Eq.~\ref{eq:mhdu}. Like before, if we make
$B=1$, we recover the known expression derived
in~\cite{bender_asymptotic_1974} for the number of multigraphs with
imposed in/out-degree sequence, obtained for the limit $k^{-/+} \ll N$.

\section{Blockmodel detection}\label{sec:detection}

The central problem which motivated large part of the existing
literature on stochastic blockmodels is the detection of the most likely
ensemble which generated a given network realization. Solving this
problem allows one to infer latent block structures in empirical data,
providing a meaningful way of grouping vertices in equivalence
classes. Blockmodel detection stands in contrast to the usual approach
of community detection~\cite{fortunato_community_2010}, which focuses
almost solely on specific block structures where nodes are connected in
dense groups, which are sparsely connected to each other (this
corresponds to the special case of a stochastic blockmodel where the
diagonal elements of the matrix $e_{rs}$ are the largest).

As mentioned in the introduction, the stochastic blockmodel entropy can
be used directly as a log-likelihood function $\mathcal{L} = \ln
\mathcal{P} = -\mathcal{S}$, if one assumes that each network
realization in the ensemble occurs with the same probability
$\mathcal{P} = 1/\Omega$. Maximizing this log-likelihood can be used as
a well-justified method of \emph{inferring} the most likely blockmodel
which generated a given network
realization~\cite{bickel_nonparametric_2009, karrer_stochastic_2011}.
Stochastic blockmodels belong to the family of exponential random
graphs~\cite{holland_exponential_1981, wasserman_logit_1996}, and as
such display the asymptotic property of consistently generating networks
from which the original model can be inferred, if the networks are large
enough~\cite{bickel_nonparametric_2009, zhao_consistency_2011}.

In~\cite{karrer_stochastic_2011} a log-likelihood function for the
degree-corrected stochastic blockmodel ensemble was derived, in the
limit where the network is sufficiently sparse. As we will show, using
entropy expressions derived here, we obtain a log-likelihood function
which generalizes the expression obtained
in~\cite{karrer_stochastic_2011}, which is recovered when one assumes
not only that the graph is sufficiently sparse, but also that the degree
distribution is not very broad. Since this specific situation has been
covered in detail in~\cite{karrer_stochastic_2011}, we focus here on a
simple, representative example where the degree distribution is broad
enough so that if this limit is assumed, it leads to misleading
results. Network topologies which exhibit entropic effects due to broad
degree distributions are often found in real systems, of which perhaps
the best-known is the internet~\cite{park_origin_2003,
johnson_entropic_2010}. We also consider the situation where there are
``extrinsic'' degree correlations, in addition to the latent block
structure. The same methods can be used in a straightforward way for
multigraph or directed ensembles, using the corresponding entropy
expressions derived in the previous sections.

Given a network realization, the task of blockmodel inference consists
in finding a block partition $\{b_i\} \in [0, B-1]^N$ of the vertices,
which maximizes the log-likelihood function $\mathcal{L}$. Considering,
for instance, the degree-corrected blockmodel ensemble with ``soft''
degree constraints~\footnote{We could easily use any of the other
entropy expressions derived previously, to accommodate the diverse
variants of the ensemble, which could be directed, mutltigraphs,
etc. However, the use of the expressions derived for the ``hard'' degree
constraints have a more limited validity, since it assumes stronger
sparsity conditions. We focus therefore on ensembles with soft degree
constraints, since they are more generally applicable.}, using
Eq.~\ref{eq:ssdu} one can write the following log-likelihood function,
\begin{widetext}
\begin{equation}\label{eq:L}
  \mathcal{L}(G|\{b_i\}) =
  \sum_{rs} e_{rs} \ln\left(\frac{e_{rs}}{e_re_s}\right)
  + \sum_{rs} n_rn_s \sum_{l=1}^{L} \frac{1}{l(l+1)}\left(\frac{e_{rs}}{e_re_s}\right)^{l+1}\left<k^{l+1}\right>_r\left<k^{l+1}\right>_s,
\end{equation}
\end{widetext}
where the terms not depending on the block partition $\{b_i\}$ were
dropped, and $L$ is a parameter which controls how many terms in the sum
are considered. Using this function, we encompass the following cases:
\begin{enumerate}
\item For $L=0$ the objective function derived
  in~\cite{karrer_stochastic_2011} is recovered, which corresponds to
  the situation where the second term can be neglected entirely.
\item For $L>0$, higher order corrections are considered, which may be
  relevant if the higher moments of the degree sequence on each block
  are sufficiently large.
\end{enumerate}

The general approach used here is to maximize $\mathcal{L}$, as given by
Eq.~\ref{eq:L}, by starting with a random block partition, and changing
the block membership of a given vertex to the value for which
$\mathcal{L}$ is maximal, and proceeding in the same way repeatedly for
all vertices, until no further improvement is possible. The algorithmic
complexity of updating the membership of a single vertex in a such a
``greedy'' manner is $O(B(B(L+1) + \avg{k}))$, which does not depend on
the system size, and therefore is efficient as long as $B$ is not too
large. However this algorithm will often get stuck in a local maximum,
so one has to start over from a different random partition, and compare
the maximum obtained. Repeating this a few times is often enough to find
the optimal solution~\footnote{This simple method can be very
  inefficient in certain cases, specially if the network is very large,
  since one may always finish in local maxima which are far away from
  the global optimum. We have also used the better variant know as the
  Kernighan-Lin algorithm~\cite{kernighan_efficient_1970}, adapted to
  the blockmodel problem in~\cite{karrer_stochastic_2011}, which can
  escape such local solutions. However, for the simple examples
  considered here, we found almost no difference in the results.}.

In the following we will consider a representative example where the
terms for $L>0$ are indeed relevant and result in different block
partitions, when compared to $L=0$. Instead of testing the general
approach in difficult cases, we deliberately choose a very simple
scenario, where the block structure is very well defined, in order to
make the block identification as easy as possible. However, as we will
see, even in these rather extreme cases, not properly accounting for the
correct entropic effects will lead to spurious results, which is the
case with $L=0$.

\subsection{Intrinsic degree correlations}\label{sec:instrinsic}

In order to illustrate the use of the objective function given by
Eq.~\ref{eq:L} we will consider a simple diagonal blockmodel defined as,
\begin{equation}\label{eq:intrinsic}
  e_{rs} \propto w\delta_{rs} + (1-w)(1-\delta_{rs}),
\end{equation}
where $w \in [0,1]$ is free parameter, and all blocks have equal
size. Furthermore, independently of the block membership, the degrees
will be distributed according to a Zipf distribution within a certain
range,
\begin{equation}
  p_k\propto
  \begin{cases}
    k^{-\gamma}, &\text{ if } k\in[k_\text{min},k_\text{max}] \\
    0, &\text{ otherwise.}
  \end{cases}
\end{equation}
This choice allows for a precise control of how broad the distribution
is. Here we will consider a typical sample from this ensemble, with
$N=10^3$ vertices, $B=4$, a strong block structure with $w = 0.99$, and
degree distribution with $\gamma=1.1$ and $[k_\text{min},k_\text{max}] =
[30, 200]$. As mentioned before, this strong block structure is
deliberately chosen to make the detection task more straightforward. The
sample was generated using the Metropolis-Hastings
algorithm~\cite{metropolis_equation_1953, hastings_monte_1970}, by
starting with some network with a degree sequence sampled from the
desired distribution, and the block labels distributed randomly among
the nodes. At each step, the end point of two randomly chosen edges are
swapped, such that the degree sequence is preserved. The probability
difference $\Delta p = p' - p$ is computed, where $p\propto
\sum_{ij}A_{ij}e_{b_i, b_j}$ is the probability of observing the given
network before the move, and $p'$ is the same probability after the
move. If $\Delta P$ is positive the move is accepted, otherwise it is
rejected with probability $1-p'/p$. Additionally, a move is always
rejected if it generates a parallel edge or a self-loop. If the
probabilities are nonzero, this defines a Markov chain which fulfills
detailed balance, and which is known to be
ergodic~\cite{mcavaney_constrained_1981, mcavaney_simple_1981,
coolen_constrained_2009}, and thus generates samples with the correct
probability after equilibrium is reached~\footnote{This algorithm
actually generates samples from the \emph{canonical} ensemble, since it
allows for fluctuations in the numbers $e_{rs}$. However, as mentioned
in Sec.~\ref{sec:def}, this ensemble is equivalent to the microcanonical
version for sufficiently large samples.}.

\begin{figure} 
  \begin{minipage}{0.49\columnwidth}
    \begin{center}
      \includegraphics[width=\columnwidth]{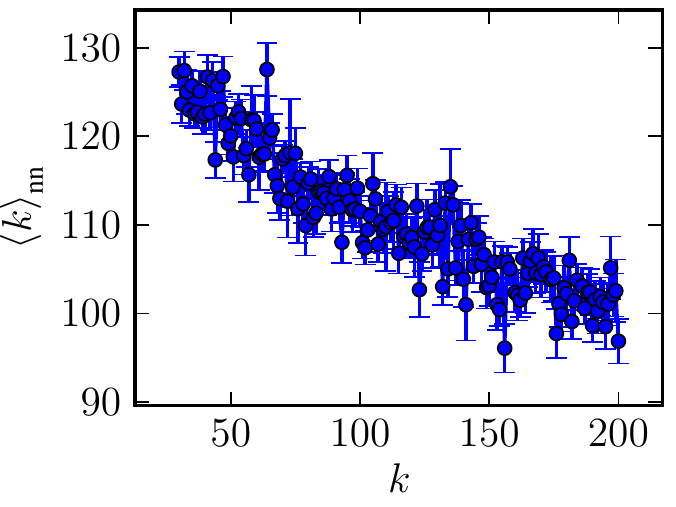}
    \end{center}
  \end{minipage}
  \begin{minipage}{0.49\columnwidth}
    \begin{center}
      \includegraphics[width=\columnwidth]{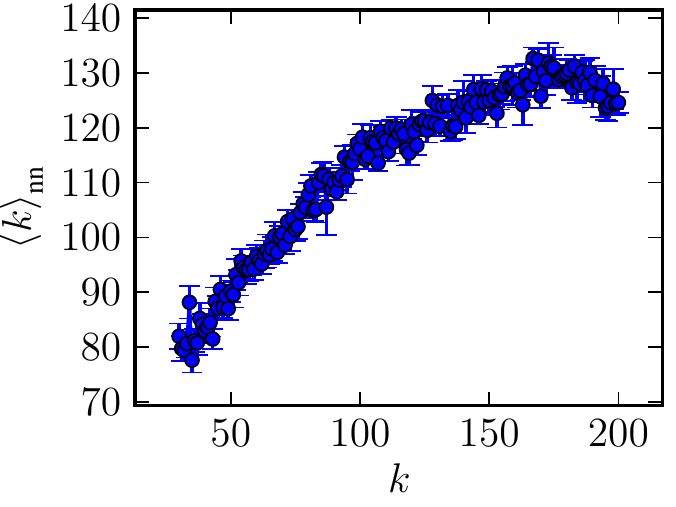}
    \end{center}
  \end{minipage}
  \caption{(Color online) Average nearest neighbour degree $\avg{k}_{\text{nn}}(k)$, as
    a function of the degree of the originating vertex $k$, for the
    model with intrinsic (left) and extrinsic (right) degree
    correlations.\label{fig:deg-corr}}
\end{figure}

As can be seen in Fig.~\ref{fig:deg-corr}, the degree distribution is
broad enough to cause intrinsic dissortative degree correlations in the
generated sample. In the following, the same single sample from the
ensemble will be used, to mimic the situation of empirically obtained
data. However, we have repeated the analysis for different samples from
the ensemble, and found always very similar results.

It is usually the case that one does not know \emph{a priori} which
value of $B$ is the most appropriate. Hence, one must obtain the best
partitions for several $B$ values, and choose the one with the largest
value of $\mathcal{L}$. However, the values of $\mathcal{L}$ will always
increase monotonically with $B$, since the number of suitable models
will become larger, while the data remains the same, culminating in the
extreme situation where each vertex will belong to its own block, and
the inferred $e_{rs}$ parameters will be given directly by the adjacency
matrix~\footnote{An alternative which circumvents this problem is the
so-called Maximum A Posteriori (MAP) approach, which uses parameter
\emph{distributions}, instead of a single set of parameters when
maximizing the log-likelihood. Instead of the log-likelihood increasing
monotonically with $B$, the parameter distributions become broader
instead. This approach has been applied to the degree-corrected
stochastic blockmodel in~\cite{reichardt_interplay_2011}, using belief
propagation. This method, however, has the disadvantage of being
numerically less efficient for large networks.}. One can estimate how
$\mathcal{L}$ should increase with $B$ by exploiting the fact the first
term in Eq.~\ref{eq:L} has the same functional form as the mutual
information of two random variables $x$, $y$,
\begin{equation}\label{eq:mi}
  I(x, y) = \sum_{xy} p_{xy}\ln\left(\frac{p_{xy}}{p_xp_y}\right),
\end{equation}
where $p_{xy}$ is the joint distribution of both variables. It is a
known fact that the mutual information calculated from empirical
distributions suffers from an upwards systematic bias which disappears
only as the number of samples goes to
infinity~\cite{treves_upward_1995}. Assuming the fluctuations of the
counts in each bin of the distribution are independent, one can
calculate this bias analytically as $\Delta I(x, y) = (X - 1)(Y -
1)/2N_s
+ O(1/N_s^2)$, where $X$ and $Y$ are the number of possible values of the
$x$ and $y$ variables, respectively, and $N_s$ is the number of
empirical samples~\cite{treves_upward_1995}. Using this information, one
can obtain an estimation for the dependence of $\mathcal{L}$ on $B$,
\begin{equation}\label{eq:bias}
  \mathcal{L}^* \approx \mathcal{L} - (B - 1)^2,
\end{equation}
where $\mathcal{L}^*$ is the expected ``true'' value of the
log-likelihood, if the sample size goes to infinity~\footnote{We note
that Eq.~\ref{eq:bias}
  should be understood only as a rule of thumb which gives a \emph{lower
    bound} on the bias of $\mathcal{L}$, since it is obtained only from
  the first term of Eq.~\ref{eq:L}, and assumes that the number of
  blocks in each partition fluctuates independently, which is not likely
  to hold in general since the block partition is a result of an
  optimization algorithm.}. This can be used to roughly differentiate
  between situations where the log-likelihood is increasing due to new
  block structures which are being discovered, and when it is only due
  to an artifact of the limited data.

\begin{figure} 
  \begin{minipage}{0.49\columnwidth}
    \begin{center}
      \includegraphics[width=\columnwidth]{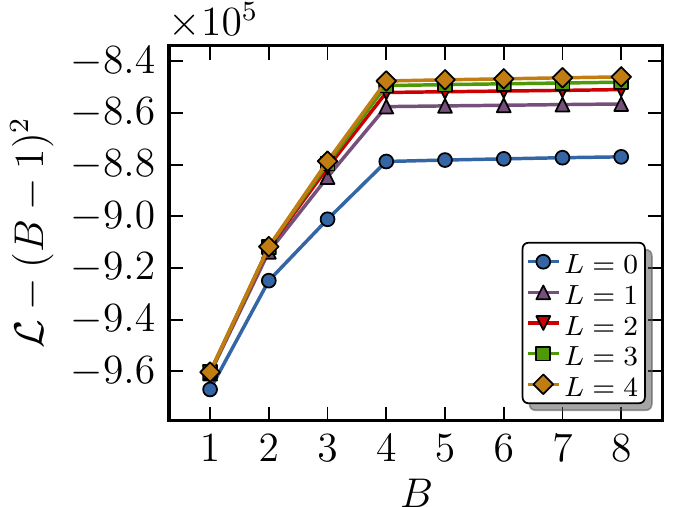}
    \end{center}
  \end{minipage}
  \begin{minipage}{0.49\columnwidth}
    \begin{center}
      \includegraphics[width=\columnwidth]{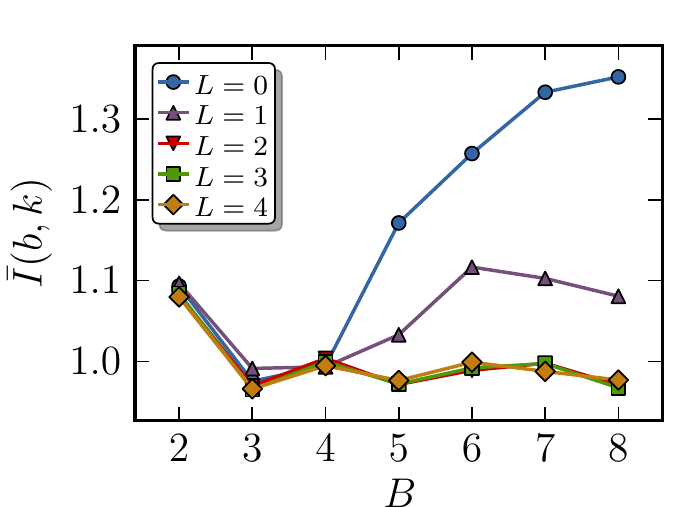}
    \end{center}
  \end{minipage} \caption{(Color online) Left: Optimized log-likelihood $\mathcal{L}$
  (Eq.~\ref{eq:L}) as a function of
    $B$, for different values of $L$, for the same sample from the
    ensemble with intrinsic degree correlations. Right: Average
    normalized mutual information (Eq.~\ref{eq:nmi}) between the degree
    sequence and the block partition, as a function of $B$, for
    different values of $L$. \label{fig:uncorr-stats}}
\end{figure}

\begin{figure} 
  \begin{minipage}{0.49\columnwidth}
    \begin{center}
      \includegraphics[width=\columnwidth]{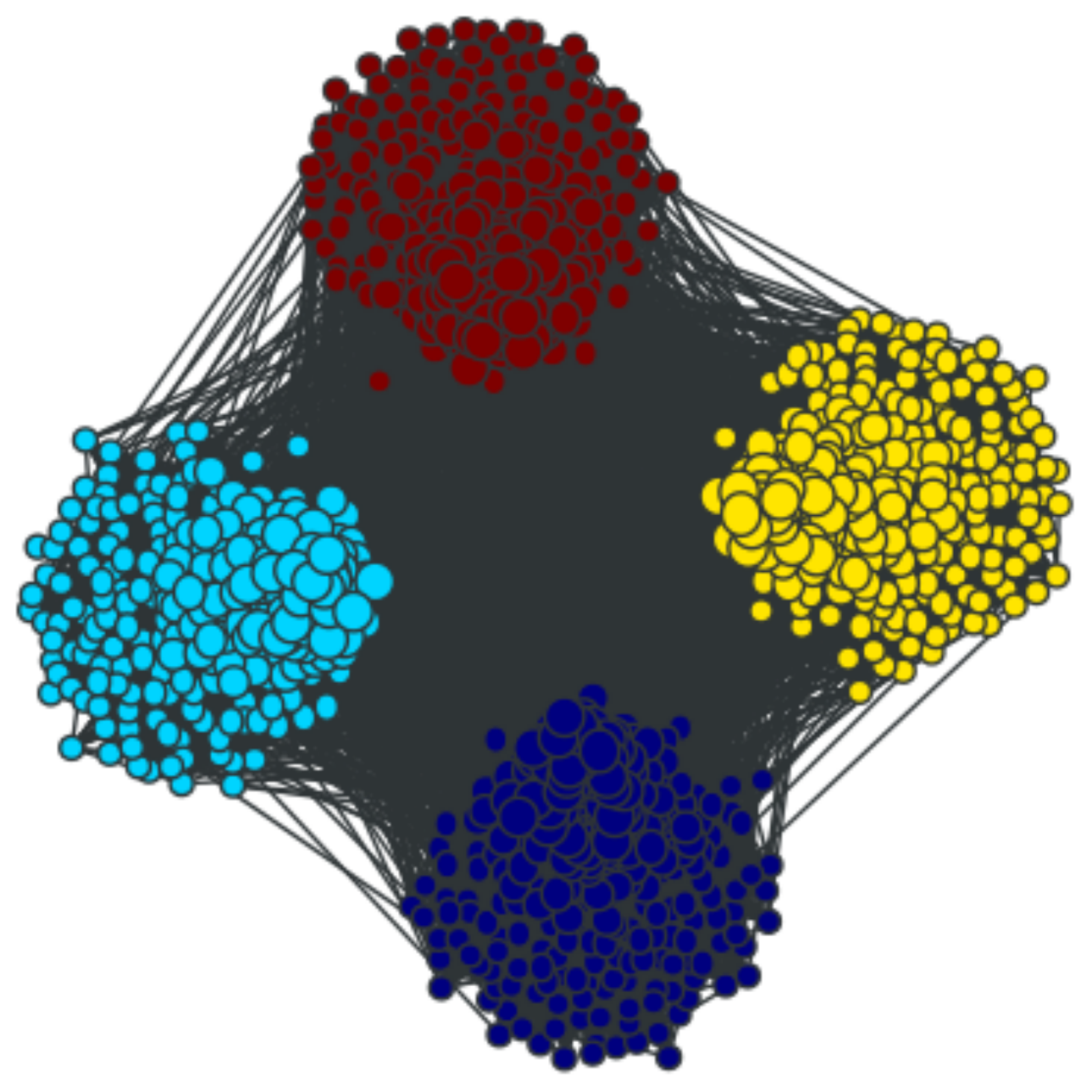}
      $B=4, L=\{0,1,2,3,4\}$
    \end{center}
  \end{minipage}
  \begin{minipage}{0.49\columnwidth}
    \begin{center}
      \includegraphics[width=\columnwidth]{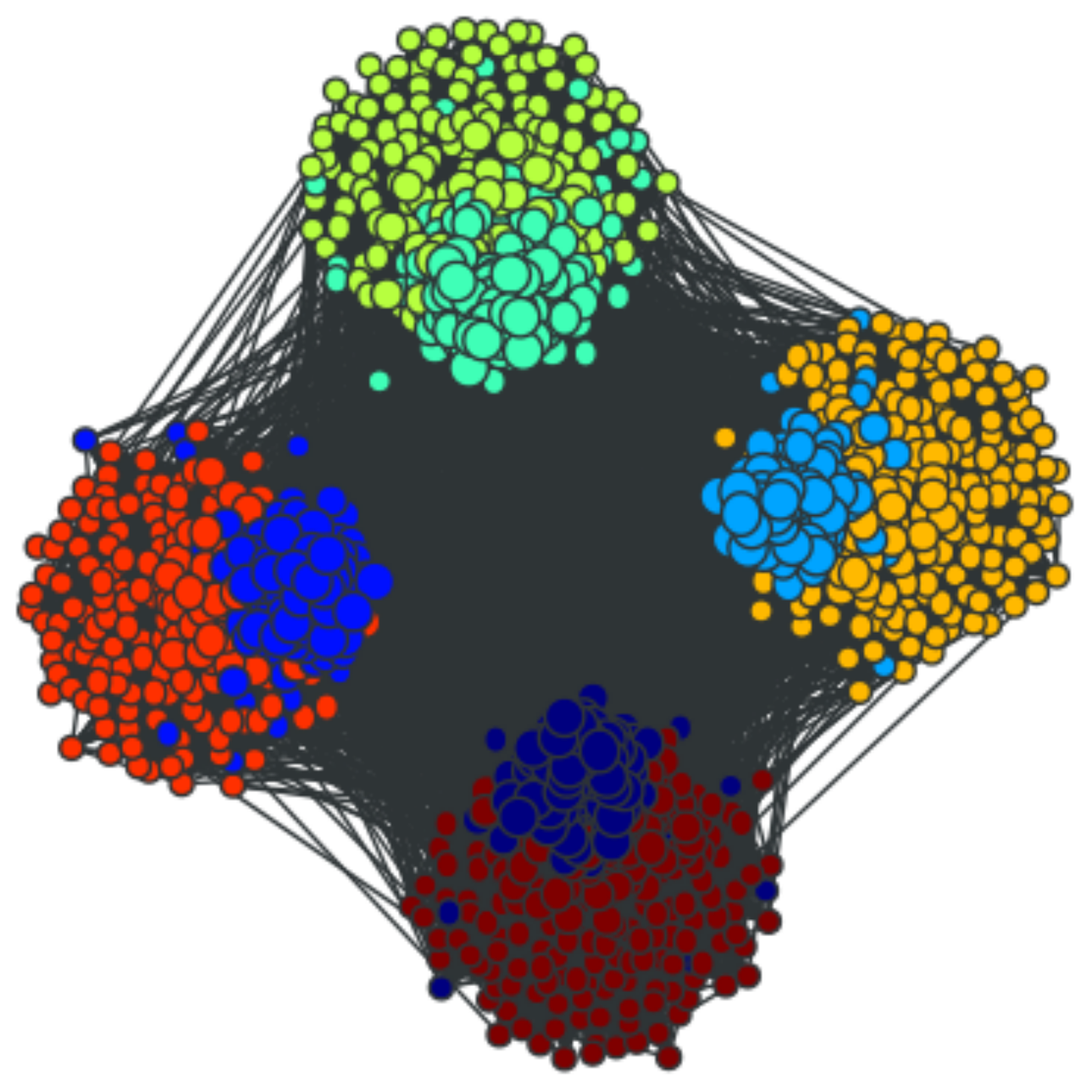}
      $B=8, L=0$
    \end{center}
  \end{minipage}
  \begin{minipage}{0.49\columnwidth}
    \begin{center}
      \includegraphics[width=\columnwidth]{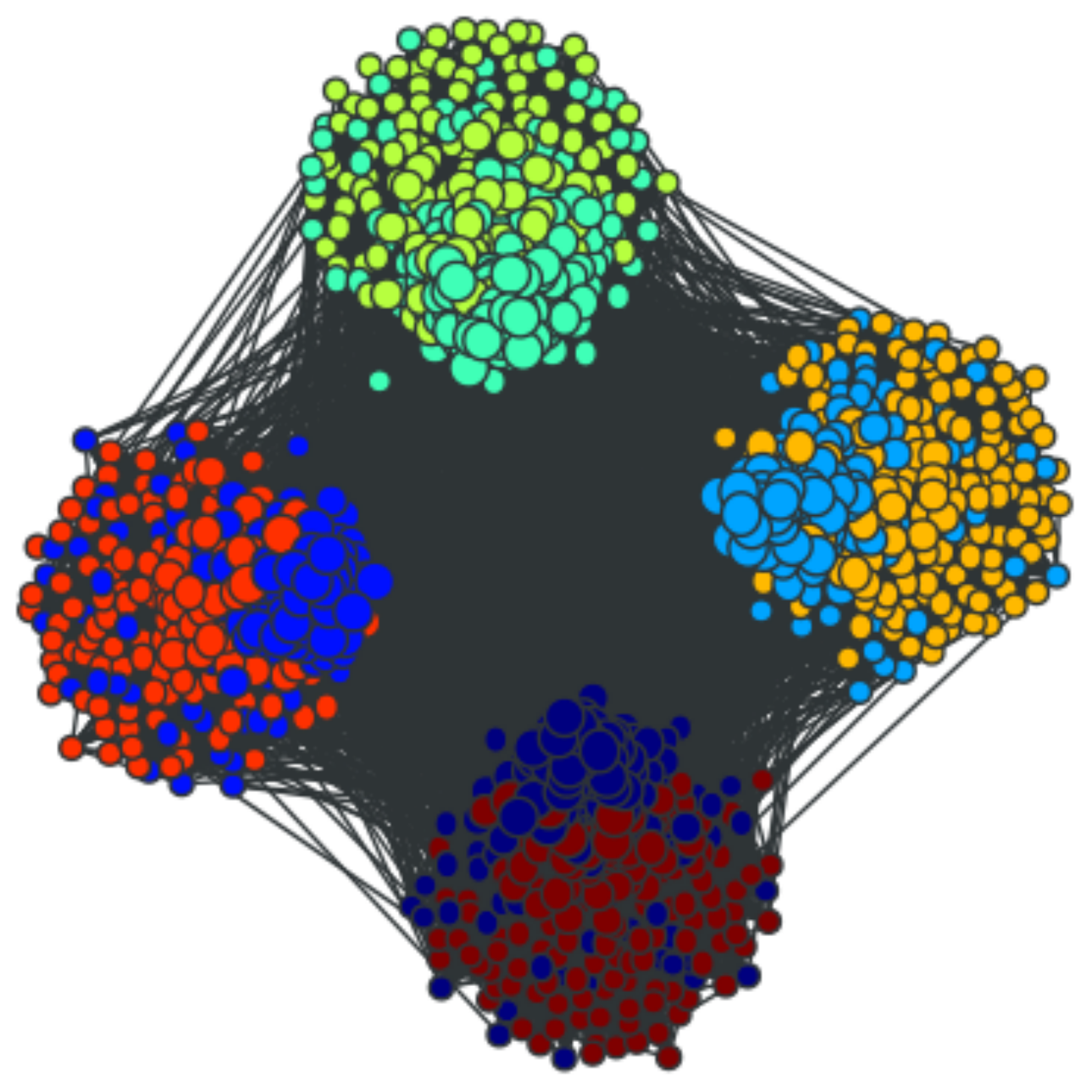}
      $B=8, L=1$
    \end{center}
  \end{minipage}
  \begin{minipage}{0.49\columnwidth}
    \begin{center}
      \includegraphics[width=\columnwidth]{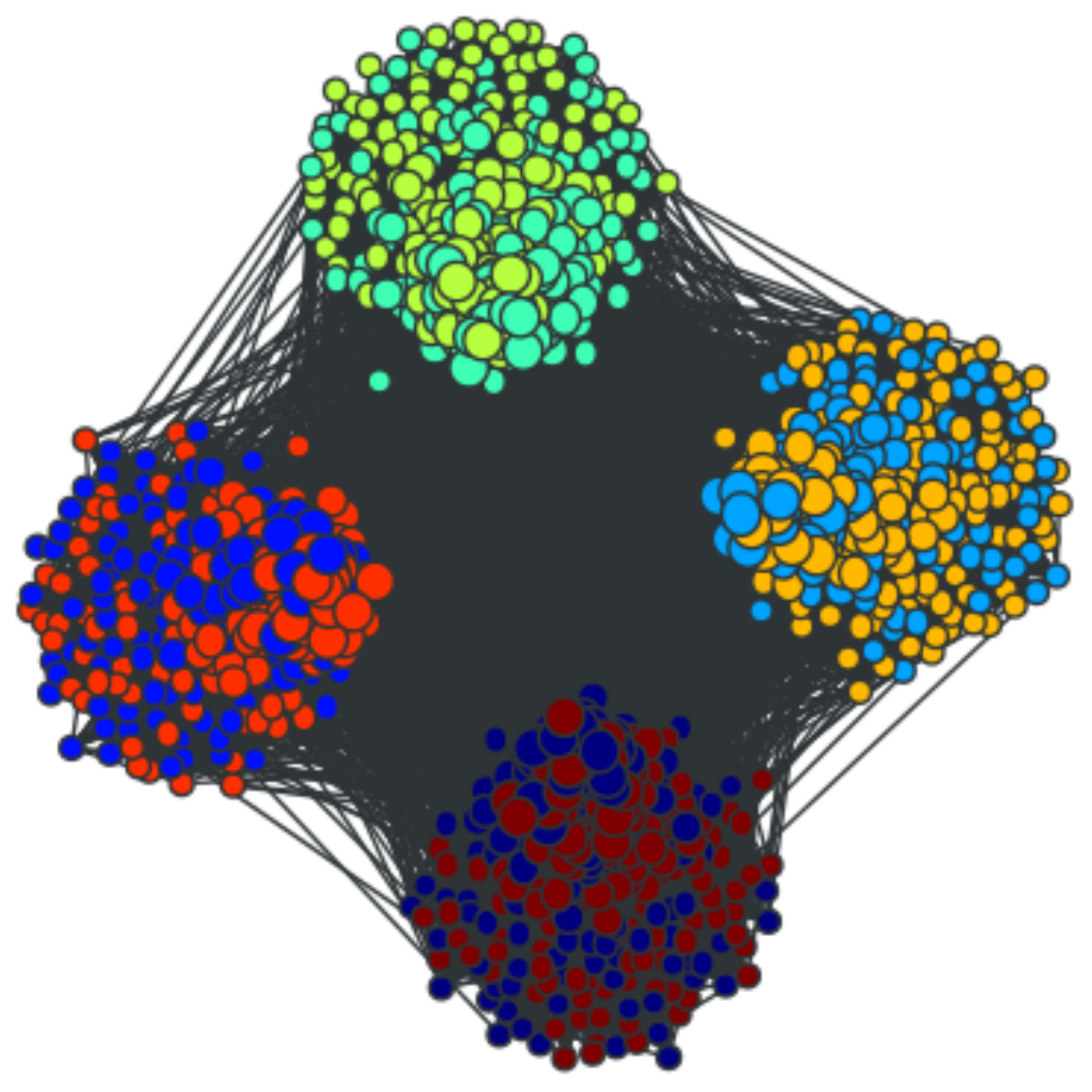}
      $B=8, L=2$
    \end{center}
  \end{minipage}
  \begin{minipage}{0.49\columnwidth}
    \begin{center}
      \includegraphics[width=\columnwidth]{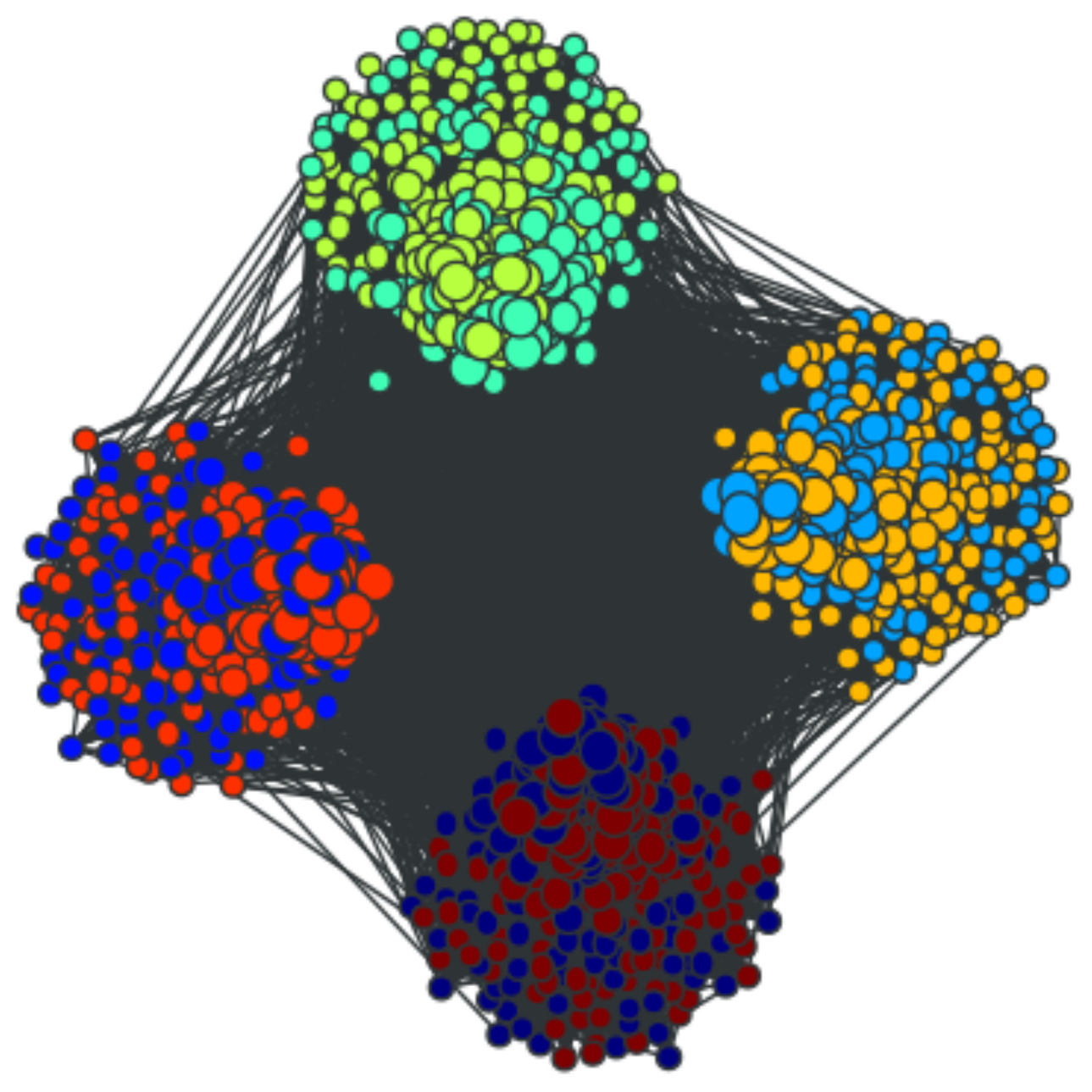}
      $B=8, L=3$
    \end{center}
  \end{minipage}
  \begin{minipage}{0.49\columnwidth}
    \begin{center}
      \includegraphics[width=\columnwidth]{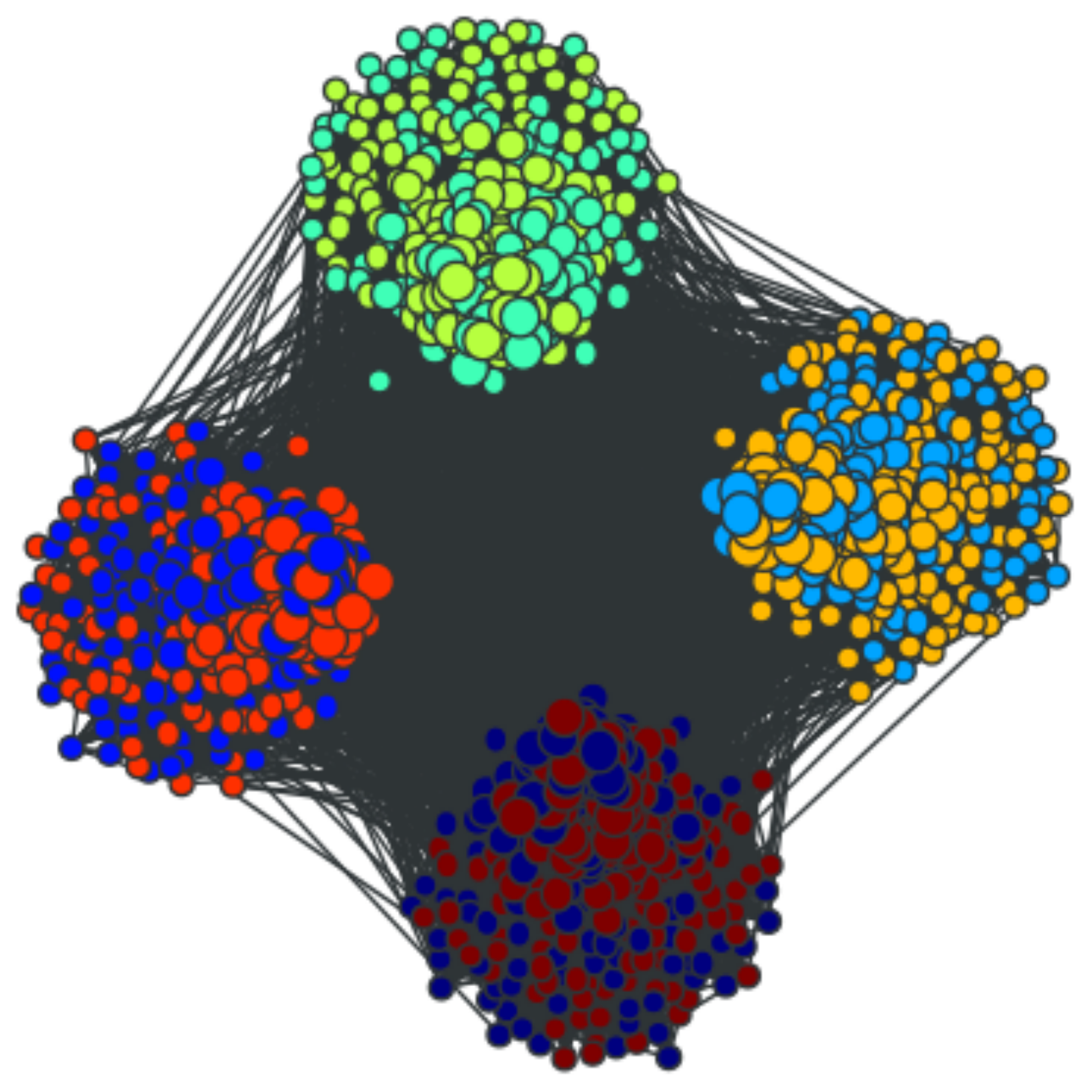}
      $B=8, L=4$
    \end{center}
  \end{minipage}
  \caption{(Color online) Obtained block partitions for different values of $B$ and
    $L$, for the same sample of the ensemble with intrinsic degree
    correlations. The colors indicate the partition, and the size of the
    vertices is proportional to the degree. Nodes of high degree are
    pushed towards the center of the layout. Note that for $B=8$ and $L
    \in [0,1]$, the nodes of high degree are segregated into separate
    blocks. \label{fig:part}}
\end{figure}

In Fig.~\ref{fig:uncorr-stats} are shown the values of $\mathcal{L}$ for
different $L$, for the same sample of the ensemble above. The likelihood
increases monotonically until $B=4$, after which it does not increase
significantly. The values of $\mathcal{L}$ are significantly different
for different $L$ (which shows that the higher order terms in
Eq.~\ref{eq:L} should indeed not be neglected), but all curves indicate
$B=4$ as being the ``true'' partition size, which is indeed
correct. However, a closer inspection of the resulting partitions
reveals important differences. In Fig.~\ref{fig:part} are shown some of
the obtained partitions for different values of $L$ and $B$. For $B=4$,
all values of $L$ result in the same partition, which corresponds
exactly to the correct partition. For larger values of $B$, however, the
obtained partitions differ a lot more than one would guess by looking at
the values of $\mathcal{L}$ alone. For $B=8$ and $L=0$ one sees a clear
division into $8$ blocks, which strongly separates vertices of different
degrees. This could easily be mistaken for a true partition, despite the
fact that it is nothing more than an entropic artifact of the broad
degree distribution. Indeed if one increases $L$, the optimal partition
becomes eventually a random sub-partition of the correct $B=4$
structure. In this particular example, $L=2$ is enough to obtain the
correct result, and the higher values result in the same partition, with
only negligible differences.

The correlation of the block partition with the degree sequence can be
computed more precisely by using the mutual information $I(b,k)$
(Eq.~\ref{eq:mi}), between the block labels and the degrees.  Since we
want to compare partitions obtained for different values of $B$, and
changing $B$ will invariably change $I(b,k)$, we use instead the average
normalized mutual information, defined here as,
\begin{equation}\label{eq:nmi}
  \bar{I}(b, k) = \left<\frac{I(b,k)}{I(r,k)}\right>,
\end{equation}
where $I(r,k)$ is the mutual information of the degree sequence and a
random block partition $\{r_i\}$, obtained by shuffling the block labels
$\{b_i\}$. The average is taken over several independent realizations of
$\{r_i\}$. If the block partition is uncorrelated with the degree
sequence, one should have that $\bar{I}(b, k)$ is close to one, since
there are no intrinsic correlations between the correct partition and
the degrees. The values of $\bar{I}(b, k)$ are shown in
Fig~\ref{fig:uncorr-stats}. One sees clearly that the results for lower
values of $L$ are significantly correlated with the degree sequence, and
that for $L\ge 2$ the correlation essentially vanishes.

The reason why the log-likelihood with $L=0$ delivers spurious block
structures is intimately related to the fact that the degree
distribution is this case is broad. This causes the remaining terms of
Eq.~\ref{eq:L} to become relevant, as they represent the entropic cost
of an edge leading to a block with a broader degree distribution. On the
other hand, the same entropic cost is responsible for the dissortative
degree correlations seen in Fig.~\ref{fig:deg-corr}. This is in fact
inconsistent with the assumption made when deriving Eq.~\ref{eq:ssdu},
namely Eq.~\ref{eq:classical_limit}, which says that there are no such
degree correlations. This is indeed true, and it means that
Eq.~\ref{eq:L}, even for $L\to\infty$, is still an approximation which
neglects certain entropic effects. However, as mentioned previously, it
still captures a large portion of the entropic cost of placing an edge
incident to a block with a broad degree sequence, and this is the reason
why it can be used to infer the correct block structure in the example
shown. The same performance should be expected in situations where the
intrinsic degree correlations are present, but not ``too strong'' as to
require better approximations. Indeed, as was discussed previously
following the derivation of Eq.~\ref{eq:ssdu}, for networks with very
large degrees it may be that Eq.~\ref{eq:L} diverges, for sufficiently
large $L$. However, this situation can be managed adequately. In
Sec.~\ref{sec:simple_soft} we computed the entropy for the ensemble with
soft degree constraints and arbitrary degree correlations, given in
Eq.~\ref{eq:ssd}. This expression is exact, and can be used as a
log-likelihood in the extreme situations where Eq.~\ref{eq:L} is not a
good approximation. The downside is that one needs to infer much more
parameters, since the model is defined by the full matrix $e_{(r,k),
(s,k)}$, which makes the maximization of $\mathcal{L}$ less
computationally efficient, and may result in overfitting. A more
efficient method will be described in the next section, which consists
in separating vertices in groups of similar degree, and using this
auxiliary partition to infer the actual block structure. This can be
done in way which allows one to control how much information needs to be
inferred, such that the degree correlations (intrinsic or otherwise)
have been sufficiently accounted for.

\subsection{Extrinsic degree correlations}\label{sec:extrinsic}

We consider now the case where there are arbitrary extrinsic degree
correlations (although the method described here also works well in
situations with strong intrinsic degree correlations which are not well
captured by Eq.~\ref{eq:L}). As an example, we will use a modified
version of the blockmodel ensemble used in the previous section, which
includes assortative degree correlations, defined as
\begin{equation}\label{eq:extrinsic}
  e_{(r,k), (s, k')} \propto \frac{e_{rs}}{1 + |k-k'|},
\end{equation}
where $e_{rs}$ is given by Eq.~\ref{eq:intrinsic}. Similarly to the
previous case, we consider a typical sample from this ensemble, with
$N=10^3$ vertices, $B=4$, a block structure with $w = 0.99$, and degree
distribution with $\gamma=1.1$ and $[k_\text{min},k_\text{max}] = [30,
200]$. The degree correlations obtained in this sample is show in
Fig.~\ref{fig:deg-corr}.

\begin{figure} 
  \begin{minipage}{0.49\columnwidth}
    \begin{center}
      \includegraphics[width=\columnwidth]{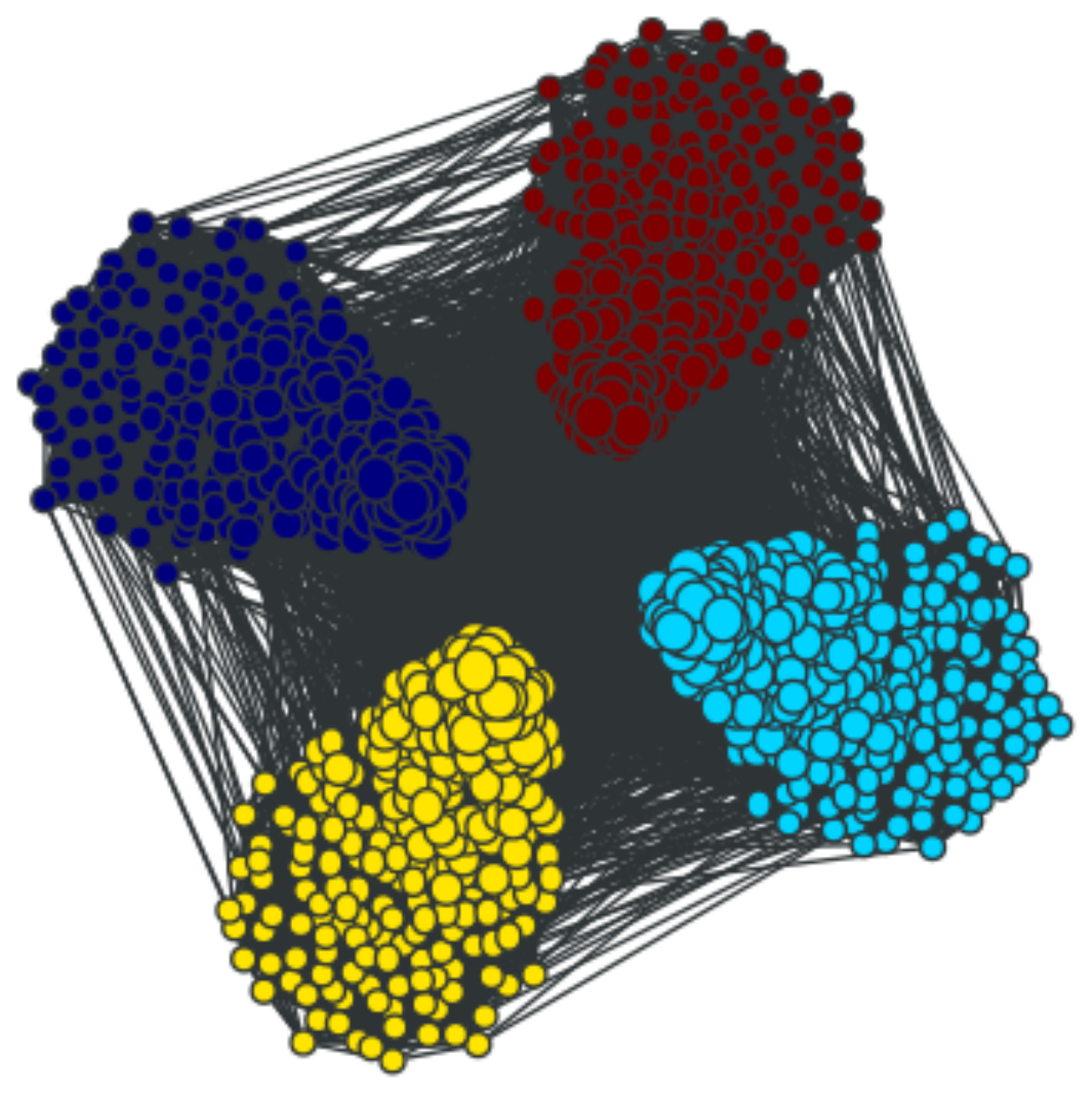}
      $B=4$
    \end{center}
  \end{minipage}
  \begin{minipage}{0.49\columnwidth}
    \begin{center}
      \includegraphics[width=\columnwidth]{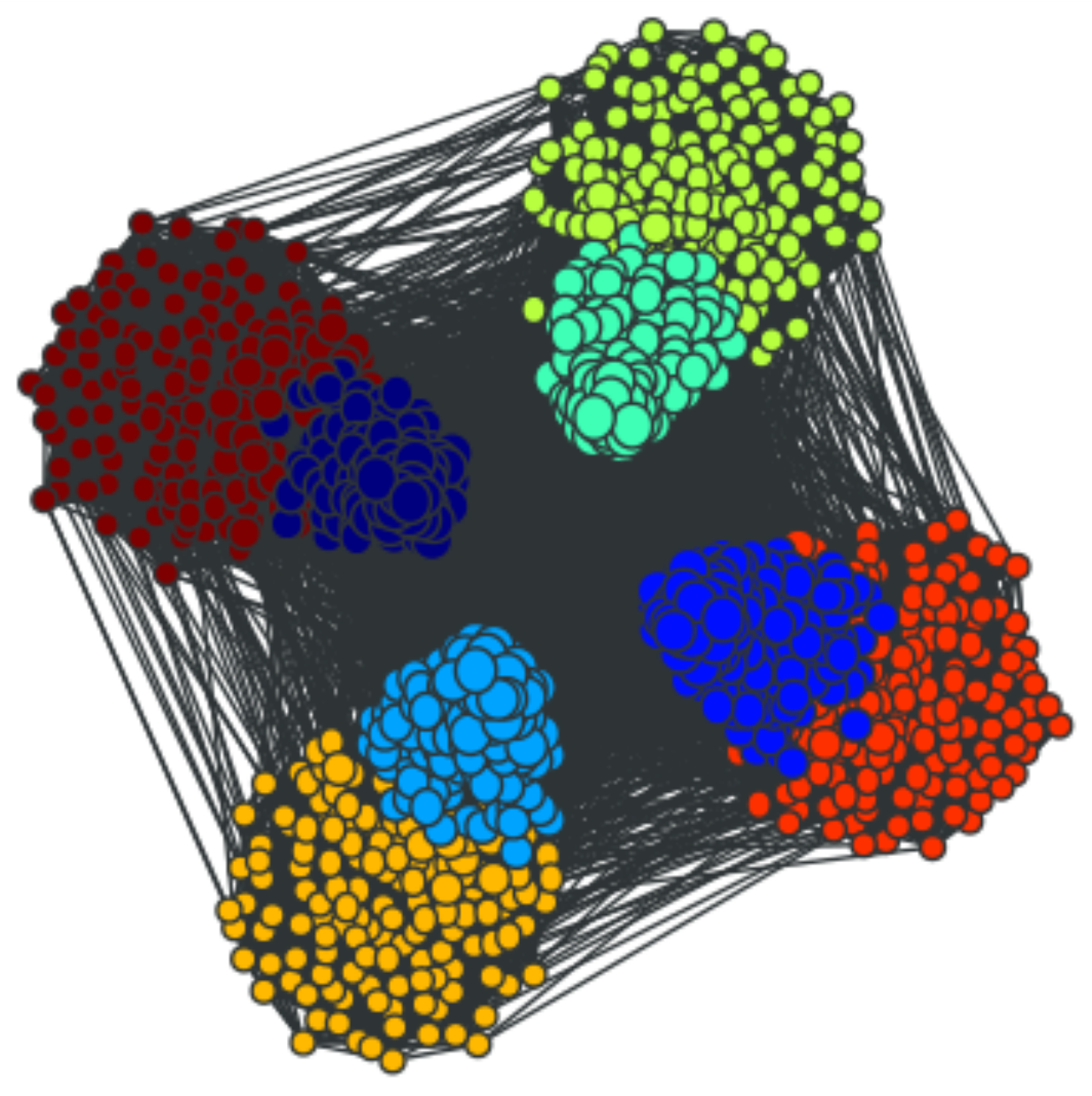}
      $B=8$
    \end{center}
  \end{minipage}
  \caption{(Color online) Inferred block partitions for the model with extrinsic degree
    correlations, obtained my maximizing the log-likelihood
    $\mathcal{L}$, given by Eq.~\ref{eq:L}, for different values of $B$
    and $L=2$.\label{fig:part-bad}}
\end{figure}

\begin{figure} 
  \begin{minipage}{0.49\columnwidth}
    \begin{center}
      \includegraphics[width=\columnwidth]{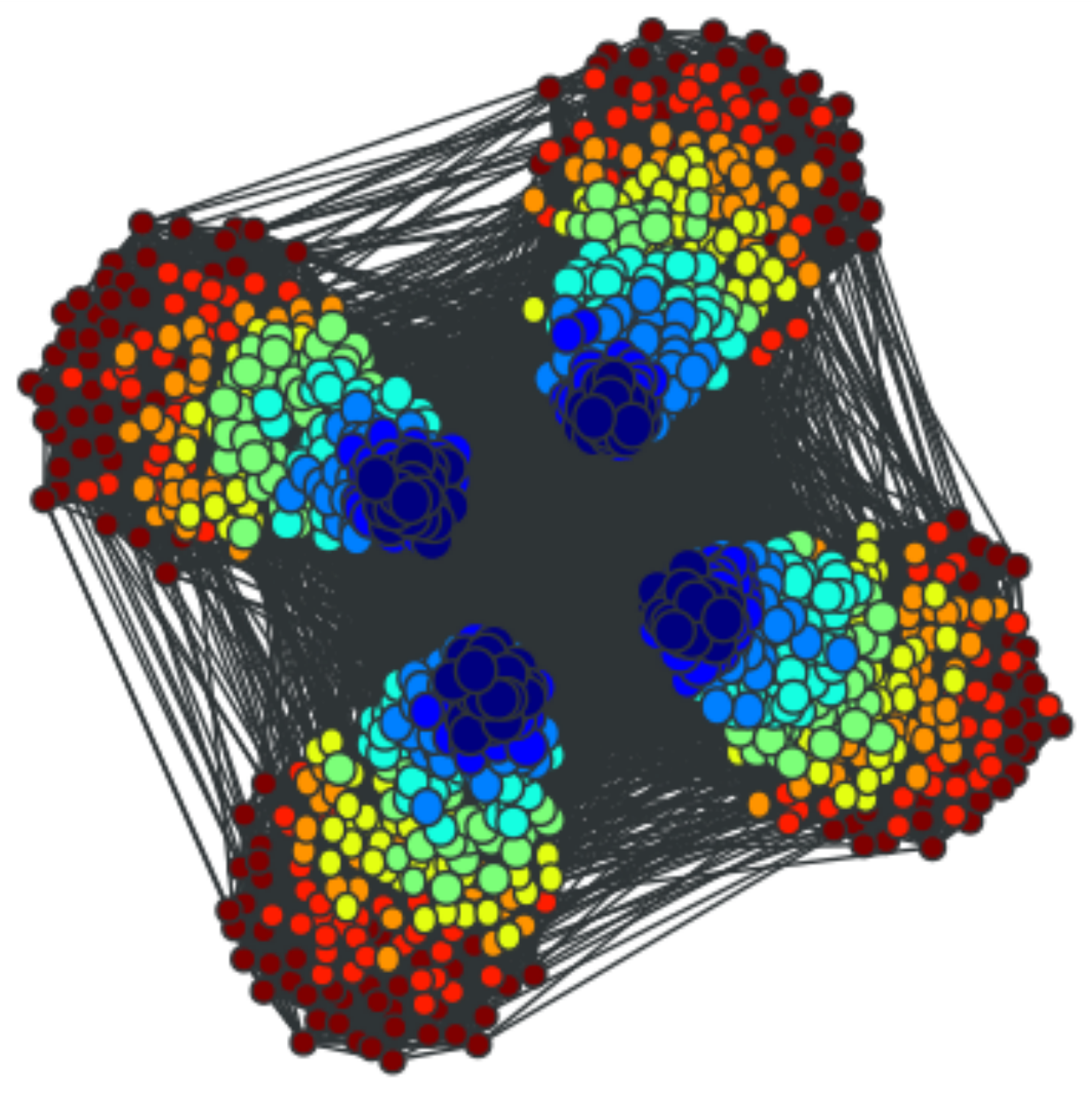}
      Auxiliary partition $\{d_i\}$, with $D=8$.
    \end{center}
  \end{minipage}
  \begin{minipage}{0.49\columnwidth}
    \begin{center}
      \includegraphics[width=\columnwidth]{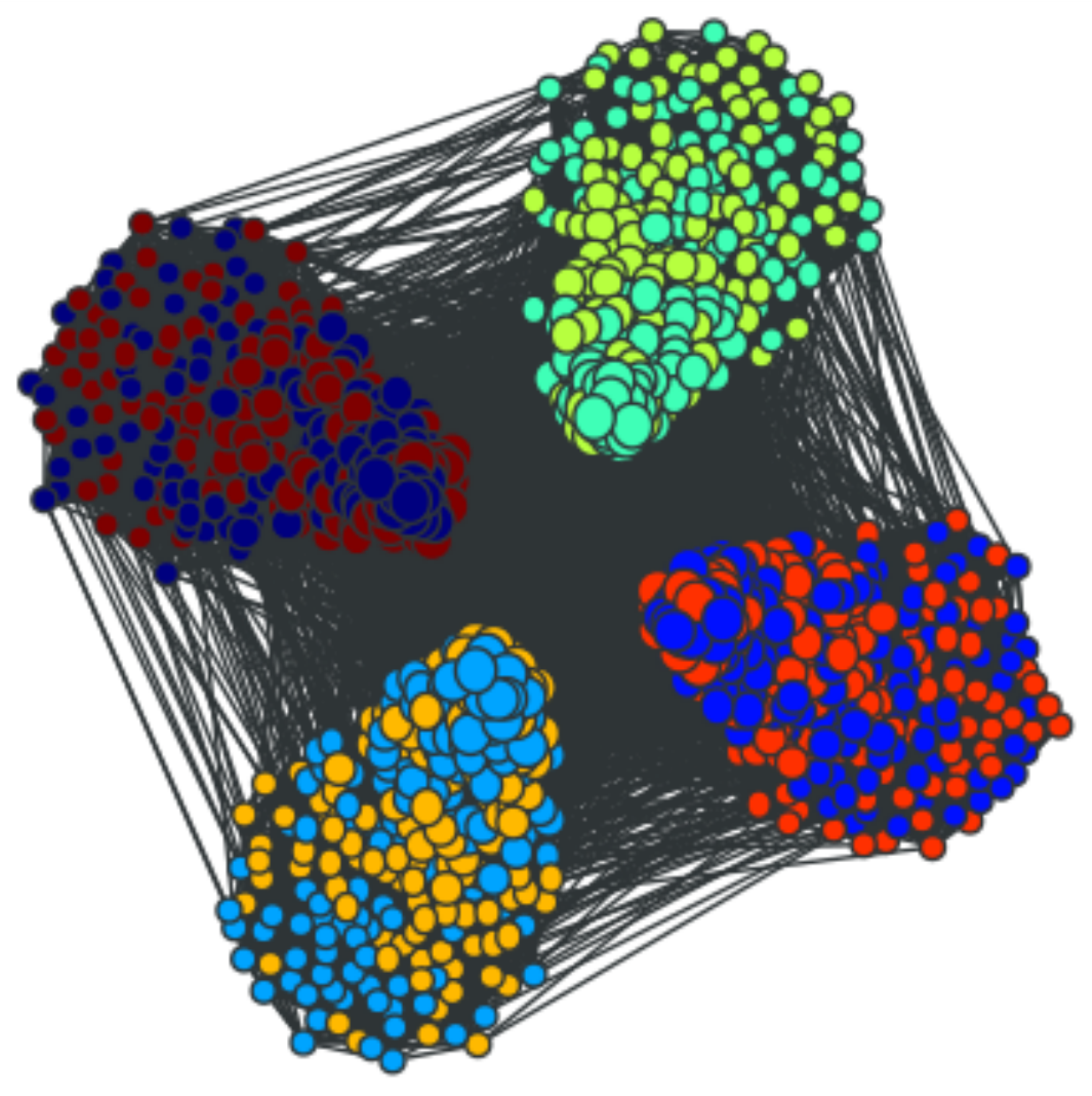}
      Inferred partition $\{b_i\}$, with $B=8$.
    \end{center}
  \end{minipage}
  \caption{(Color online) Auxiliary and inferred block partitions for a sample of the
    ensemble with intrinsic degree correlations.\label{fig:aux-part}}
\end{figure}

\begin{figure} 
  \begin{minipage}{0.49\columnwidth}
    \begin{center}
      \includegraphics[width=\columnwidth]{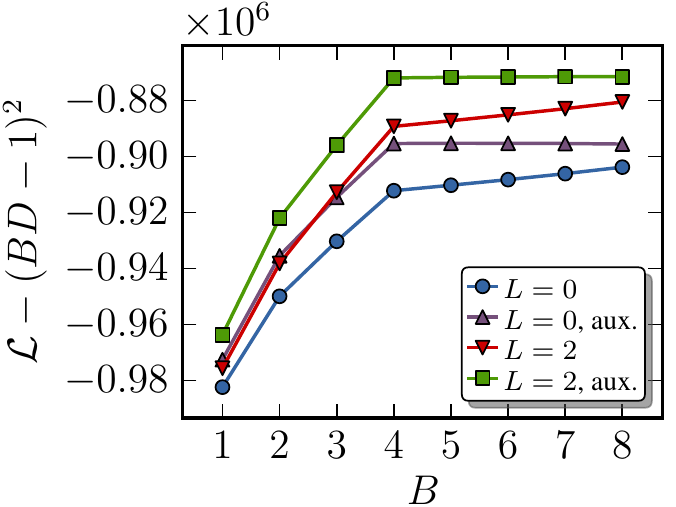}
    \end{center}
  \end{minipage}
  \begin{minipage}{0.49\columnwidth}
    \begin{center}
      \includegraphics[width=\columnwidth]{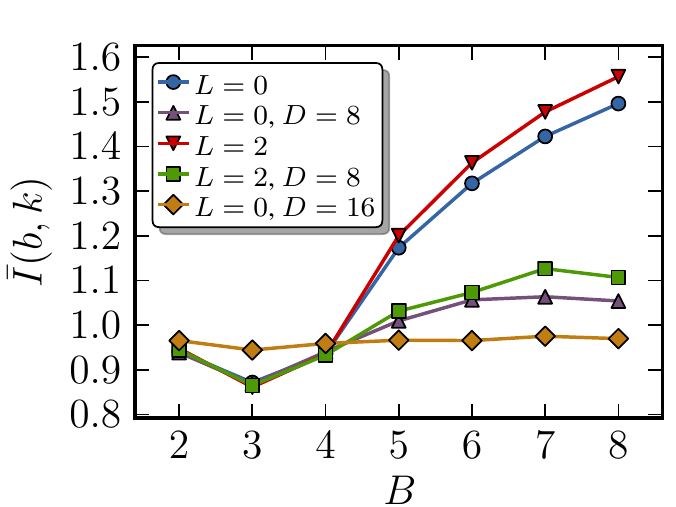}
    \end{center}
  \end{minipage}
  \caption{(Color online)
    Left: Optimized log-likelihood $\mathcal{L}$ (Eq.~\ref{eq:L}) as a
    function of $B$, for different values of $L$, for the same sample
    from the ensemble with extrinsic degree correlations. The legend
    ``aux.'' indicates results obtained with the auxiliary degree-based
    partition described in the text. Right: Average normalized mutual
    information (Eq.~\ref{eq:nmi}) between the degree sequence and the
    block partition, as a function of $B$, for different values of $L$,
    and size of the auxiliary partition $D$ (or without it if $D$ is
    omitted). \label{fig:corr-stats-aux}}
\end{figure}

If one does not know, or ignores, that there are degree correlations
present, and attempts to detect the most likely block structure using
Eq.~\ref{eq:L}, one obtains block partitions shown in
Fig.~\ref{fig:part-bad}. Due to the high segregation of the modules, one
indeed finds the correct block partition for $B=4$, but as the value of
$B$ is increased, one finds increasingly many ``sub-blocks''
corresponding to groups vertices of different degrees. This is simply a
manifestation of the degree correlations present in
Eq.~\ref{eq:extrinsic}. As Fig.~\ref{fig:part-bad} shows, the
log-likelihood increases steadily with larger $B$ values, indicating
that the ``true'' block structure has not yet been found. Indeed one
would need to make $B\sim 4K$, where $K$ is the number of different
degrees in the network, to finally capture the complete structure. The
correct inferred partition in this case would put vertices of the same
degree in their own block, which we can label as $(r,k)$. In this
situation, Eq.~\ref{eq:ssdu} becomes no longer an approximation, since
Eq.~\ref{eq:sparse_limit} will also hold exactly, and it becomes
identical to Eq.~\ref{eq:ssd}, which we could use instead as a
log-likelihood (which effectively removes the parameter $L$).  Strictly
speaking, Eq.~\ref{eq:ssd} is entirely sufficient to detect any block
structure with arbitrary degree correlations, either intrinsic or
extrinsic. In practice, however, it is cumbersome to use since it
requires the inference a large amount of parameters, namely the full
$e_{(r,k),(r,k)}$ matrix of size $(BK)^2$ (of which half the elements
are independent parameters), as well as the $n_{(r,k)}$ vector of size
$BK$. The number of different degrees $K$ is often significantly
large. For the specific example shown in Fig.~\ref{fig:part-bad} we have
$K=168$, which results in a parameter matrix which is much larger than
the number of edges in the network. This is an undesired situation,
since with such a large number of parameters, not only it becomes easier
to get trapped in local maxima when optimizing $\mathcal{L}$, but also
it becomes impossible to discern between actual features of the inferred
model and stochastic fluctuations which are frozen in the network
structure. However, it is possible to circumvent this problem using the
following approach. Before attempting to infer the block partition
$\{b_i\}$, one constructs an auxiliary partition $\{d_i\}$ which remains
fixed throughout the entire process. The auxiliary partition separates
vertices in $D$ blocks representing degree bins, so that vertices in the
same block have similar degrees. Exactly how large should be each degree
block, and how the bin boundaries should be chosen will depend in
general of specific network properties; however a good starting point is
to separate them into bins such that the total number of bins $D$ is as
small as possible, while at the same time keeping the degree variance
within each bin also small. Furthermore, one should also avoid having
degree bins with very few vertices, since this is more likely to lead to
artifacts due to lack of statistics. With this auxiliary partition in
hand, one can proceed to infer a block partition $\{b_i\}$ into $B$
blocks, such that the combined block label of a given vertex $i$ is
$(b_i,d_i)$. The log-likelihood is computed using Eq.~\ref{eq:L}, using
the full $(b,d)$ block labels to differentiate between blocks. If the
$\{d_i\}$ partition is reasonably chosen, the degree correlations will
be inferred automatically, and from the $\{b_i\}$ partition it is
possible to extract the block structure which is independent from degree
correlations. Note however that after this procedure the $b_i$ labels
\emph{by themselves} do not represent a meaningful partition, since any
relabeling of the form $(r, d)\leftrightarrow (s, d)$, for the same
value of $d$, results in an entirely equivalent block structure. In
order to obtain a meaningful $\{b_i\}$ partition, it is necessary to
proceed as follows,
\begin{enumerate}
\item Maximize $\mathcal{L}$ using auxiliary the partition, $\{d_i\}$,
  obtaining the best partition $\{(b_i, d_i)\}$.
\item Swap labels $(r, d) \leftrightarrow (s, d)$, within the same
      auxiliary block $d$, such that the
  log-likelihood $\mathcal{L}$, \emph{ignoring the auxiliary partition}
  $\{d_i\}$, is maximized.
\end{enumerate}
In step 2, the labels are swapped until no further improvement is
possible. After step 2 is completed, the blockmodel obtained in step 1
remains unchanged, but the block labels $\{b_i\}$ now have a clear
meaning, since they represent the best overall block structure, ignoring
the auxiliary partition, among the possibilities which are equivalent to
the inferred block partition.

In the left of Fig.~\ref{fig:aux-part} is shown an example auxiliary
partition, with $D=8$, and bin widths chosen so that all groups have
approximately the same size. On the right is shown the inferred
$\{b_i\}$ partition with $B=8$, using the auxiliary partition, after the
label swap step described above. Notice how the correlations with degree
can no longer be distinguished visually. Observing how the
log-likelihood increases with $B$ (see Fig.~\ref{fig:corr-stats-aux}),
the results with the auxiliary partition point more convincingly to the
$B=4$ structure, since it does not increase significantly for increasing
block numbers.  Fig~\ref{fig:corr-stats-aux} also shows the average
normalized mutual information between the block partitions and the
degrees, and indeed the difference between the inference with and
without the block partition is significant. For $D=8$ one can still
measure a residual correlation, but by increasing the auxiliary
partition to $D=16$ virtually removes it, which is still significantly
smaller than the total number of degrees $K=168$.

\section{Conclusion}\label{sec:conclusion}

We have calculated analytical expressions for the entropy of stochastic
blockmodel ensembles, both in its traditional and degree-corrected
forms. We have considered all the fundamental variants of the ensembles,
including directed and undirected graphs, as well as degree sequences
implemented as soft and hard constraints. The expressions derived
represent generalizations of the known entropies of random graphs with
arbitrary degree sequence~\cite{bender_asymptotic_1974,
  bender_asymptotic_1978, wormald_models_1999, bianconi_entropy_2009},
  which are easily recovered by setting the number of blocks to one.

As a straightforward application of the derived entropy functions, we
applied them to the task of blockmodel inference, given observed
data. We showed that this method can be used even in situations where
there are intrinsic (i.e. with an entropic origin) degree correlations,
and can be easily adapted to the case with arbitrary extrinsic degree
correlations. This approach represents a generalization of the one
presented in~\cite{karrer_stochastic_2011}, which is only expected to
work well with sparse graphs without very broad degree sequences.

Furthermore, the blockmodel entropy could also be used as a more refined
method to infer the relevance of topological features in empirical
networks~\cite{bianconi_assessing_2009}, and to determine the
statistical significance of modular network
partitions~\cite{lancichinetti_statistical_2010,
radicchi_combinatorial_2010, lancichinetti_finding_2011}.

Beyond the task of block detection, the knowledge of the entropy of
these ensembles can be used to directly obtain the equilibrium
properties of network systems which possess an energy function which
depends directly on the block structure. Indeed this has been used
in~\cite{peixoto_emergence_2012} to construct a simplified model of gene
regulatory system, in which the robustness can be expressed in terms of
the block structure, functioning as an energy function. The evolutionary
process acting on the system was mapped to a Gibbs ensemble, where the
selective pressure plays the role of temperature. The equilibrium
properties were obtained by minimizing the free energy, which was
written using the blockmodel entropy. This model in particular exhibited
a topological phase transition at higher values of selective pressure,
where the network becomes assembled in a core-periphery structure, which
is very similar to what is observed in real gene networks. We speculate
that the blockmodel entropy can be used in the same manner to obtain
properties of wide variety of adaptive
networks~\cite{gross_adaptive_2008}, for which stochastic blockmodels
are adequate models.

\bibliographystyle{apsrev4-1}
\bibliography{bib}

\begin{thebibliography}{78}%
\makeatletter
\providecommand \@ifxundefined [1]{%
 \@ifx{#1\undefined}
}%
\providecommand \@ifnum [1]{%
 \ifnum #1\expandafter \@firstoftwo
 \else \expandafter \@secondoftwo
 \fi
}%
\providecommand \@ifx [1]{%
 \ifx #1\expandafter \@firstoftwo
 \else \expandafter \@secondoftwo
 \fi
}%
\providecommand \natexlab [1]{#1}%
\providecommand \enquote  [1]{``#1''}%
\providecommand \bibnamefont  [1]{#1}%
\providecommand \bibfnamefont [1]{#1}%
\providecommand \citenamefont [1]{#1}%
\providecommand \href@noop [0]{\@secondoftwo}%
\providecommand \href [0]{\begingroup \@sanitize@url \@href}%
\providecommand \@href[1]{\@@startlink{#1}\@@href}%
\providecommand \@@href[1]{\endgroup#1\@@endlink}%
\providecommand \@sanitize@url [0]{\catcode `\\12\catcode `\$12\catcode
  `\&12\catcode `\#12\catcode `\^12\catcode `\_12\catcode `\%12\relax}%
\providecommand \@@startlink[1]{}%
\providecommand \@@endlink[0]{}%
\providecommand \url  [0]{\begingroup\@sanitize@url \@url }%
\providecommand \@url [1]{\endgroup\@href {#1}{\urlprefix }}%
\providecommand \urlprefix  [0]{URL }%
\providecommand \Eprint [0]{\href }%
\providecommand \doibase [0]{http://dx.doi.org/}%
\providecommand \selectlanguage [0]{\@gobble}%
\providecommand \bibinfo  [0]{\@secondoftwo}%
\providecommand \bibfield  [0]{\@secondoftwo}%
\providecommand \translation [1]{[#1]}%
\providecommand \BibitemOpen [0]{}%
\providecommand \bibitemStop [0]{}%
\providecommand \bibitemNoStop [0]{.\EOS\space}%
\providecommand \EOS [0]{\spacefactor3000\relax}%
\providecommand \BibitemShut  [1]{\csname bibitem#1\endcsname}%
\let\auto@bib@innerbib\@empty
\bibitem [{\citenamefont {Holland}\ \emph {et~al.}(1983)\citenamefont
  {Holland}, \citenamefont {Laskey},\ and\ \citenamefont
  {Leinhardt}}]{holland_stochastic_1983}%
  \BibitemOpen
  \bibfield  {author} {\bibinfo {author} {\bibfnamefont {P.~W.}\ \bibnamefont
  {Holland}}, \bibinfo {author} {\bibfnamefont {K.~B.}\ \bibnamefont {Laskey}},
  \ and\ \bibinfo {author} {\bibfnamefont {S.}~\bibnamefont {Leinhardt}},\
  }\href {\doibase 16/0378-8733(83)90021-7} {\bibfield  {journal} {\bibinfo
  {journal} {Social Networks}\ }\textbf {\bibinfo {volume} {5}},\ \bibinfo
  {pages} {109} (\bibinfo {year} {1983})}\BibitemShut {NoStop}%
\bibitem [{\citenamefont {Fienberg}\ \emph {et~al.}(1985)\citenamefont
  {Fienberg}, \citenamefont {Meyer},\ and\ \citenamefont
  {Wasserman}}]{fienberg_statistical_1985}%
  \BibitemOpen
  \bibfield  {author} {\bibinfo {author} {\bibfnamefont {S.~E.}\ \bibnamefont
  {Fienberg}}, \bibinfo {author} {\bibfnamefont {M.~M.}\ \bibnamefont {Meyer}},
  \ and\ \bibinfo {author} {\bibfnamefont {S.~S.}\ \bibnamefont {Wasserman}},\
  }\href {\doibase 10.2307/2288040} {\bibfield  {journal} {\bibinfo  {journal}
  {Journal of the American Statistical Association}\ }\textbf {\bibinfo
  {volume} {80}},\ \bibinfo {pages} {51} (\bibinfo {year} {1985})}\BibitemShut
  {NoStop}%
\bibitem [{\citenamefont {Faust}\ and\ \citenamefont
  {Wasserman}(1992)}]{faust_blockmodels:_1992}%
  \BibitemOpen
  \bibfield  {author} {\bibinfo {author} {\bibfnamefont {K.}~\bibnamefont
  {Faust}}\ and\ \bibinfo {author} {\bibfnamefont {S.}~\bibnamefont
  {Wasserman}},\ }\href {\doibase 16/0378-8733(92)90013-W} {\bibfield
  {journal} {\bibinfo  {journal} {Social Networks}\ }\textbf {\bibinfo {volume}
  {14}},\ \bibinfo {pages} {5} (\bibinfo {year} {1992})}\BibitemShut {NoStop}%
\bibitem [{\citenamefont {Anderson}\ \emph {et~al.}(1992)\citenamefont
  {Anderson}, \citenamefont {Wasserman},\ and\ \citenamefont
  {Faust}}]{anderson_building_1992}%
  \BibitemOpen
  \bibfield  {author} {\bibinfo {author} {\bibfnamefont {C.~J.}\ \bibnamefont
  {Anderson}}, \bibinfo {author} {\bibfnamefont {S.}~\bibnamefont {Wasserman}},
  \ and\ \bibinfo {author} {\bibfnamefont {K.}~\bibnamefont {Faust}},\ }\href
  {\doibase 16/0378-8733(92)90017-2} {\bibfield  {journal} {\bibinfo  {journal}
  {Social Networks}\ }\textbf {\bibinfo {volume} {14}},\ \bibinfo {pages} {137}
  (\bibinfo {year} {1992})}\BibitemShut {NoStop}%
\bibitem [{\citenamefont {Doreian}\ \emph {et~al.}(2004)\citenamefont
  {Doreian}, \citenamefont {Batagelj},\ and\ \citenamefont
  {Ferligoj}}]{doreian_generalized_2004}%
  \BibitemOpen
  \bibfield  {author} {\bibinfo {author} {\bibfnamefont {P.}~\bibnamefont
  {Doreian}}, \bibinfo {author} {\bibfnamefont {V.}~\bibnamefont {Batagelj}}, \
  and\ \bibinfo {author} {\bibfnamefont {A.}~\bibnamefont {Ferligoj}},\
  }\href@noop {} {\emph {\bibinfo {title} {Generalized Blockmodeling}}}\
  (\bibinfo  {publisher} {Cambridge University Press},\ \bibinfo {year}
  {2004})\BibitemShut {NoStop}%
\bibitem [{\citenamefont {Brandes}\ \emph {et~al.}(2010)\citenamefont
  {Brandes}, \citenamefont {Lerner},\ and\ \citenamefont
  {Nagel}}]{brandes_network_2010}%
  \BibitemOpen
  \bibfield  {author} {\bibinfo {author} {\bibfnamefont {U.}~\bibnamefont
  {Brandes}}, \bibinfo {author} {\bibfnamefont {J.}~\bibnamefont {Lerner}}, \
  and\ \bibinfo {author} {\bibfnamefont {U.}~\bibnamefont {Nagel}},\ }\href
  {\doibase 10.1007/s11634-010-0074-3} {\bibfield  {journal} {\bibinfo
  {journal} {Advances in Data Analysis and Classification}\ }\textbf {\bibinfo
  {volume} {5}},\ \bibinfo {pages} {81} (\bibinfo {year} {2010})}\BibitemShut
  {NoStop}%
\bibitem [{\citenamefont {Fortunato}(2010)}]{fortunato_community_2010}%
  \BibitemOpen
  \bibfield  {author} {\bibinfo {author} {\bibfnamefont {S.}~\bibnamefont
  {Fortunato}},\ }\href {\doibase 16/j.physrep.2009.11.002} {\bibfield
  {journal} {\bibinfo  {journal} {Physics Reports}\ }\textbf {\bibinfo {volume}
  {486}},\ \bibinfo {pages} {75} (\bibinfo {year} {2010})}\BibitemShut
  {NoStop}%
\bibitem [{\citenamefont {Newman}\ and\ \citenamefont
  {Leicht}(2007)}]{newman_mixture_2007}%
  \BibitemOpen
  \bibfield  {author} {\bibinfo {author} {\bibfnamefont {M.~E.~J.}\
  \bibnamefont {Newman}}\ and\ \bibinfo {author} {\bibfnamefont {E.~A.}\
  \bibnamefont {Leicht}},\ }\href {\doibase 10.1073/pnas.0610537104} {\bibfield
   {journal} {\bibinfo  {journal} {Proceedings of the National Academy of
  Sciences}\ }\textbf {\bibinfo {volume} {104}},\ \bibinfo {pages} {9564 }
  (\bibinfo {year} {2007})}\BibitemShut {NoStop}%
\bibitem [{\citenamefont {Reichardt}\ and\ \citenamefont
  {White}(2007)}]{reichardt_role_2007}%
  \BibitemOpen
  \bibfield  {author} {\bibinfo {author} {\bibfnamefont {J.}~\bibnamefont
  {Reichardt}}\ and\ \bibinfo {author} {\bibfnamefont {D.~R.}\ \bibnamefont
  {White}},\ }\href {\doibase 10.1140/epjb/e2007-00340-y} {\bibfield  {journal}
  {\bibinfo  {journal} {The European Physical Journal B}\ }\textbf {\bibinfo
  {volume} {60}},\ \bibinfo {pages} {217} (\bibinfo {year} {2007})}\BibitemShut
  {NoStop}%
\bibitem [{\citenamefont {Bickel}\ and\ \citenamefont
  {Chen}(2009)}]{bickel_nonparametric_2009}%
  \BibitemOpen
  \bibfield  {author} {\bibinfo {author} {\bibfnamefont {P.~J.}\ \bibnamefont
  {Bickel}}\ and\ \bibinfo {author} {\bibfnamefont {A.}~\bibnamefont {Chen}},\
  }\href {\doibase 10.1073/pnas.0907096106} {\bibfield  {journal} {\bibinfo
  {journal} {Proceedings of the National Academy of Sciences}\ }\textbf
  {\bibinfo {volume} {106}},\ \bibinfo {pages} {21068 } (\bibinfo {year}
  {2009})}\BibitemShut {NoStop}%
\bibitem [{\citenamefont {Guimerà}\ and\ \citenamefont
  {{Sales-Pardo}}(2009)}]{guimera_missing_2009}%
  \BibitemOpen
  \bibfield  {author} {\bibinfo {author} {\bibfnamefont {R.}~\bibnamefont
  {Guimerà}}\ and\ \bibinfo {author} {\bibfnamefont {M.}~\bibnamefont
  {{Sales-Pardo}}},\ }\href {\doibase 10.1073/pnas.0908366106} {\bibfield
  {journal} {\bibinfo  {journal} {Proceedings of the National Academy of
  Sciences}\ }\textbf {\bibinfo {volume} {106}},\ \bibinfo {pages} {22073 }
  (\bibinfo {year} {2009})}\BibitemShut {NoStop}%
\bibitem [{\citenamefont {Karrer}\ and\ \citenamefont
  {Newman}(2011)}]{karrer_stochastic_2011}%
  \BibitemOpen
  \bibfield  {author} {\bibinfo {author} {\bibfnamefont {B.}~\bibnamefont
  {Karrer}}\ and\ \bibinfo {author} {\bibfnamefont {M.~E.~J.}\ \bibnamefont
  {Newman}},\ }\href {\doibase 10.1103/PhysRevE.83.016107} {\bibfield
  {journal} {\bibinfo  {journal} {Physical Review E}\ }\textbf {\bibinfo
  {volume} {83}},\ \bibinfo {pages} {016107} (\bibinfo {year}
  {2011})}\BibitemShut {NoStop}%
\bibitem [{\citenamefont {Ball}\ \emph {et~al.}(2011)\citenamefont {Ball},
  \citenamefont {Karrer},\ and\ \citenamefont {Newman}}]{ball_efficient_2011}%
  \BibitemOpen
  \bibfield  {author} {\bibinfo {author} {\bibfnamefont {B.}~\bibnamefont
  {Ball}}, \bibinfo {author} {\bibfnamefont {B.}~\bibnamefont {Karrer}}, \ and\
  \bibinfo {author} {\bibfnamefont {M.~E.~J.}\ \bibnamefont {Newman}},\ }\href
  {\doibase 10.1103/PhysRevE.84.036103} {\bibfield  {journal} {\bibinfo
  {journal} {Physical Review E}\ }\textbf {\bibinfo {volume} {84}},\ \bibinfo
  {pages} {036103} (\bibinfo {year} {2011})}\BibitemShut {NoStop}%
\bibitem [{\citenamefont {Reichardt}\ \emph {et~al.}(2011)\citenamefont
  {Reichardt}, \citenamefont {Alamino},\ and\ \citenamefont
  {Saad}}]{reichardt_interplay_2011}%
  \BibitemOpen
  \bibfield  {author} {\bibinfo {author} {\bibfnamefont {J.}~\bibnamefont
  {Reichardt}}, \bibinfo {author} {\bibfnamefont {R.}~\bibnamefont {Alamino}},
  \ and\ \bibinfo {author} {\bibfnamefont {D.}~\bibnamefont {Saad}},\ }\href
  {\doibase 10.1371/journal.pone.0021282} {\bibfield  {journal} {\bibinfo
  {journal} {{PLoS} {ONE}}\ }\textbf {\bibinfo {volume} {6}},\ \bibinfo {pages}
  {e21282} (\bibinfo {year} {2011})}\BibitemShut {NoStop}%
\bibitem [{\citenamefont {Decelle}\ \emph {et~al.}(2011)\citenamefont
  {Decelle}, \citenamefont {Krzakala}, \citenamefont {Moore},\ and\
  \citenamefont {Zdeborová}}]{decelle_asymptotic_2011}%
  \BibitemOpen
  \bibfield  {author} {\bibinfo {author} {\bibfnamefont {A.}~\bibnamefont
  {Decelle}}, \bibinfo {author} {\bibfnamefont {F.}~\bibnamefont {Krzakala}},
  \bibinfo {author} {\bibfnamefont {C.}~\bibnamefont {Moore}}, \ and\ \bibinfo
  {author} {\bibfnamefont {L.}~\bibnamefont {Zdeborová}},\ }\href {\doibase
  10.1103/PhysRevE.84.066106} {\bibfield  {journal} {\bibinfo  {journal}
  {Physical Review E}\ }\textbf {\bibinfo {volume} {84}},\ \bibinfo {pages}
  {066106} (\bibinfo {year} {2011})}\BibitemShut {NoStop}%
\bibitem [{\citenamefont
  {Peixoto}(2012{\natexlab{a}})}]{peixoto_emergence_2012}%
  \BibitemOpen
  \bibfield  {author} {\bibinfo {author} {\bibfnamefont {T.~P.}\ \bibnamefont
  {Peixoto}},\ }\href {\doibase 10.1103/PhysRevE.85.041908} {\bibfield
  {journal} {\bibinfo  {journal} {Physical Review E}\ }\textbf {\bibinfo
  {volume} {85}},\ \bibinfo {pages} {041908} (\bibinfo {year}
  {2012}{\natexlab{a}})}\BibitemShut {NoStop}%
\bibitem [{\citenamefont
  {Peixoto}(2012{\natexlab{b}})}]{peixoto_behavior_2012}%
  \BibitemOpen
  \bibfield  {author} {\bibinfo {author} {\bibfnamefont {T.~P.}\ \bibnamefont
  {Peixoto}},\ }\href {\doibase 10.1088/1742-5468/2012/01/P01006} {\bibfield
  {journal} {\bibinfo  {journal} {Journal of Statistical Mechanics: Theory and
  Experiment}\ }\textbf {\bibinfo {volume} {2012}},\ \bibinfo {pages} {P01006}
  (\bibinfo {year} {2012}{\natexlab{b}})}\BibitemShut {NoStop}%
\bibitem [{\citenamefont {Gross}\ and\ \citenamefont
  {Blasius}(2008)}]{gross_adaptive_2008}%
  \BibitemOpen
  \bibfield  {author} {\bibinfo {author} {\bibfnamefont {T.}~\bibnamefont
  {Gross}}\ and\ \bibinfo {author} {\bibfnamefont {B.}~\bibnamefont
  {Blasius}},\ }\href {\doibase 10.1098/rsif.2007.1229} {\bibfield  {journal}
  {\bibinfo  {journal} {Journal of The Royal Society Interface}\ }\textbf
  {\bibinfo {volume} {5}},\ \bibinfo {pages} {259 } (\bibinfo {year}
  {2008})}\BibitemShut {NoStop}%
\bibitem [{\citenamefont {Newman}(2010)}]{newman_networks:_2010}%
  \BibitemOpen
  \bibfield  {author} {\bibinfo {author} {\bibfnamefont {M.}~\bibnamefont
  {Newman}},\ }\href@noop {} {\emph {\bibinfo {title} {Networks: An
  Introduction}}}\ (\bibinfo  {publisher} {Oxford University Press},\ \bibinfo
  {year} {2010})\BibitemShut {NoStop}%
\bibitem [{\citenamefont {Newman}(2003{\natexlab{a}})}]{newman_mixing_2003}%
  \BibitemOpen
  \bibfield  {author} {\bibinfo {author} {\bibfnamefont {M.~E.~J.}\
  \bibnamefont {Newman}},\ }\href
  {http://link.aps.org/abstract/PRE/v67/e026126} {\bibfield  {journal}
  {\bibinfo  {journal} {Phys. Rev. E}\ }\textbf {\bibinfo {volume} {67}},\
  \bibinfo {pages} {026126} (\bibinfo {year} {2003}{\natexlab{a}})}\BibitemShut
  {NoStop}%
\bibitem [{\citenamefont {Girvan}\ and\ \citenamefont
  {Newman}(2002)}]{girvan_community_2002}%
  \BibitemOpen
  \bibfield  {author} {\bibinfo {author} {\bibfnamefont {M.}~\bibnamefont
  {Girvan}}\ and\ \bibinfo {author} {\bibfnamefont {M.~E.~J.}\ \bibnamefont
  {Newman}},\ }\href {\doibase 10.1073/pnas.122653799} {\bibfield  {journal}
  {\bibinfo  {journal} {Proceedings of the National Academy of Sciences}\
  }\textbf {\bibinfo {volume} {99}},\ \bibinfo {pages} {7821 } (\bibinfo {year}
  {2002})}\BibitemShut {NoStop}%
\bibitem [{\citenamefont {Bianconi}(2008)}]{bianconi_entropy_2008}%
  \BibitemOpen
  \bibfield  {author} {\bibinfo {author} {\bibfnamefont {G.}~\bibnamefont
  {Bianconi}},\ }\href {\doibase 10.1209/0295-5075/81/28005} {\bibfield
  {journal} {\bibinfo  {journal} {{EPL} {(Europhysics} Letters)}\ }\textbf
  {\bibinfo {volume} {81}},\ \bibinfo {pages} {28005} (\bibinfo {year}
  {2008})}\BibitemShut {NoStop}%
\bibitem [{\citenamefont {Anand}\ and\ \citenamefont
  {Bianconi}(2009)}]{anand_entropy_2009}%
  \BibitemOpen
  \bibfield  {author} {\bibinfo {author} {\bibfnamefont {K.}~\bibnamefont
  {Anand}}\ and\ \bibinfo {author} {\bibfnamefont {G.}~\bibnamefont
  {Bianconi}},\ }\href {\doibase 10.1103/PhysRevE.80.045102} {\bibfield
  {journal} {\bibinfo  {journal} {Physical Review E}\ }\textbf {\bibinfo
  {volume} {80}},\ \bibinfo {pages} {045102} (\bibinfo {year}
  {2009})}\BibitemShut {NoStop}%
\bibitem [{\citenamefont {Bianconi}(2009)}]{bianconi_entropy_2009}%
  \BibitemOpen
  \bibfield  {author} {\bibinfo {author} {\bibfnamefont {G.}~\bibnamefont
  {Bianconi}},\ }\href {\doibase 10.1103/PhysRevE.79.036114} {\bibfield
  {journal} {\bibinfo  {journal} {Physical Review E}\ }\textbf {\bibinfo
  {volume} {79}},\ \bibinfo {pages} {036114} (\bibinfo {year}
  {2009})}\BibitemShut {NoStop}%
\bibitem [{\citenamefont {Anand}\ and\ \citenamefont
  {Bianconi}(2010)}]{anand_gibbs_2010}%
  \BibitemOpen
  \bibfield  {author} {\bibinfo {author} {\bibfnamefont {K.}~\bibnamefont
  {Anand}}\ and\ \bibinfo {author} {\bibfnamefont {G.}~\bibnamefont
  {Bianconi}},\ }\href {\doibase 10.1103/PhysRevE.82.011116} {\bibfield
  {journal} {\bibinfo  {journal} {Physical Review E}\ }\textbf {\bibinfo
  {volume} {82}},\ \bibinfo {pages} {011116} (\bibinfo {year}
  {2010})}\BibitemShut {NoStop}%
\bibitem [{\citenamefont {Castellano}\ \emph {et~al.}(2009)\citenamefont
  {Castellano}, \citenamefont {Fortunato},\ and\ \citenamefont
  {Loreto}}]{castellano_statistical_2009}%
  \BibitemOpen
  \bibfield  {author} {\bibinfo {author} {\bibfnamefont {C.}~\bibnamefont
  {Castellano}}, \bibinfo {author} {\bibfnamefont {S.}~\bibnamefont
  {Fortunato}}, \ and\ \bibinfo {author} {\bibfnamefont {V.}~\bibnamefont
  {Loreto}},\ }\href {\doibase 10.1103/RevModPhys.81.591} {\bibfield  {journal}
  {\bibinfo  {journal} {Reviews of Modern Physics}\ }\textbf {\bibinfo {volume}
  {81}},\ \bibinfo {pages} {591} (\bibinfo {year} {2009})}\BibitemShut
  {NoStop}%
\bibitem [{\citenamefont {Strauss}(1986)}]{strauss_general_1986}%
  \BibitemOpen
  \bibfield  {author} {\bibinfo {author} {\bibfnamefont {D.}~\bibnamefont
  {Strauss}},\ }\href {\doibase 10.1137/1028156} {\bibfield  {journal}
  {\bibinfo  {journal} {{SIAM} Review}\ }\textbf {\bibinfo {volume} {28}},\
  \bibinfo {pages} {513} (\bibinfo {year} {1986})}\BibitemShut {NoStop}%
\bibitem [{\citenamefont {Burda}\ \emph {et~al.}(2001)\citenamefont {Burda},
  \citenamefont {Correia},\ and\ \citenamefont
  {Krzywicki}}]{burda_statistical_2001}%
  \BibitemOpen
  \bibfield  {author} {\bibinfo {author} {\bibfnamefont {Z.}~\bibnamefont
  {Burda}}, \bibinfo {author} {\bibfnamefont {J.~D.}\ \bibnamefont {Correia}},
  \ and\ \bibinfo {author} {\bibfnamefont {A.}~\bibnamefont {Krzywicki}},\
  }\href {\doibase 10.1103/PhysRevE.64.046118} {\bibfield  {journal} {\bibinfo
  {journal} {Physical Review E}\ }\textbf {\bibinfo {volume} {64}},\ \bibinfo
  {pages} {046118} (\bibinfo {year} {2001})}\BibitemShut {NoStop}%
\bibitem [{\citenamefont {Park}\ and\ \citenamefont
  {Newman}(2004{\natexlab{a}})}]{park_statistical_2004}%
  \BibitemOpen
  \bibfield  {author} {\bibinfo {author} {\bibfnamefont {J.}~\bibnamefont
  {Park}}\ and\ \bibinfo {author} {\bibfnamefont {M.~E.~J.}\ \bibnamefont
  {Newman}},\ }\href {\doibase 10.1103/PhysRevE.70.066117} {\bibfield
  {journal} {\bibinfo  {journal} {Physical Review E}\ }\textbf {\bibinfo
  {volume} {70}},\ \bibinfo {pages} {066117} (\bibinfo {year}
  {2004}{\natexlab{a}})}\BibitemShut {NoStop}%
\bibitem [{\citenamefont {Park}\ and\ \citenamefont
  {Newman}(2004{\natexlab{b}})}]{park_solution_2004}%
  \BibitemOpen
  \bibfield  {author} {\bibinfo {author} {\bibfnamefont {J.}~\bibnamefont
  {Park}}\ and\ \bibinfo {author} {\bibfnamefont {M.~E.~J.}\ \bibnamefont
  {Newman}},\ }\href {\doibase 10.1103/PhysRevE.70.066146} {\bibfield
  {journal} {\bibinfo  {journal} {Physical Review E}\ }\textbf {\bibinfo
  {volume} {70}},\ \bibinfo {pages} {066146} (\bibinfo {year}
  {2004}{\natexlab{b}})}\BibitemShut {NoStop}%
\bibitem [{\citenamefont {Palla}\ \emph {et~al.}(2004)\citenamefont {Palla},
  \citenamefont {Derényi}, \citenamefont {Farkas},\ and\ \citenamefont
  {Vicsek}}]{palla_statistical_2004}%
  \BibitemOpen
  \bibfield  {author} {\bibinfo {author} {\bibfnamefont {G.}~\bibnamefont
  {Palla}}, \bibinfo {author} {\bibfnamefont {I.}~\bibnamefont {Derényi}},
  \bibinfo {author} {\bibfnamefont {I.}~\bibnamefont {Farkas}}, \ and\ \bibinfo
  {author} {\bibfnamefont {T.}~\bibnamefont {Vicsek}},\ }\href {\doibase
  10.1103/PhysRevE.69.046117} {\bibfield  {journal} {\bibinfo  {journal}
  {Physical Review E}\ }\textbf {\bibinfo {volume} {69}},\ \bibinfo {pages}
  {046117} (\bibinfo {year} {2004})}\BibitemShut {NoStop}%
\bibitem [{\citenamefont {Burda}\ \emph {et~al.}(2004)\citenamefont {Burda},
  \citenamefont {Jurkiewicz},\ and\ \citenamefont
  {Krzywicki}}]{burda_network_2004}%
  \BibitemOpen
  \bibfield  {author} {\bibinfo {author} {\bibfnamefont {Z.}~\bibnamefont
  {Burda}}, \bibinfo {author} {\bibfnamefont {J.}~\bibnamefont {Jurkiewicz}}, \
  and\ \bibinfo {author} {\bibfnamefont {A.}~\bibnamefont {Krzywicki}},\ }\href
  {\doibase 10.1103/PhysRevE.69.026106} {\bibfield  {journal} {\bibinfo
  {journal} {Physical Review E}\ }\textbf {\bibinfo {volume} {69}},\ \bibinfo
  {pages} {026106} (\bibinfo {year} {2004})}\BibitemShut {NoStop}%
\bibitem [{\citenamefont {Biely}\ and\ \citenamefont
  {Thurner}(2006)}]{biely_statistical_2006}%
  \BibitemOpen
  \bibfield  {author} {\bibinfo {author} {\bibfnamefont {C.}~\bibnamefont
  {Biely}}\ and\ \bibinfo {author} {\bibfnamefont {S.}~\bibnamefont
  {Thurner}},\ }\href {\doibase 10.1103/PhysRevE.74.066116} {\bibfield
  {journal} {\bibinfo  {journal} {Physical Review E}\ }\textbf {\bibinfo
  {volume} {74}},\ \bibinfo {pages} {066116} (\bibinfo {year}
  {2006})}\BibitemShut {NoStop}%
\bibitem [{\citenamefont {Fronczak}\ \emph {et~al.}(2006)\citenamefont
  {Fronczak}, \citenamefont {Fronczak},\ and\ \citenamefont
  {Hołyst}}]{fronczak_fluctuation-dissipation_2006}%
  \BibitemOpen
  \bibfield  {author} {\bibinfo {author} {\bibfnamefont {A.}~\bibnamefont
  {Fronczak}}, \bibinfo {author} {\bibfnamefont {P.}~\bibnamefont {Fronczak}},
  \ and\ \bibinfo {author} {\bibfnamefont {J.~A.}\ \bibnamefont {Hołyst}},\
  }\href {\doibase 10.1103/PhysRevE.73.016108} {\bibfield  {journal} {\bibinfo
  {journal} {Physical Review E}\ }\textbf {\bibinfo {volume} {73}},\ \bibinfo
  {pages} {016108} (\bibinfo {year} {2006})}\BibitemShut {NoStop}%
\bibitem [{\citenamefont {Fronczak}\ \emph {et~al.}(2007)\citenamefont
  {Fronczak}, \citenamefont {Fronczak},\ and\ \citenamefont
  {Hołyst}}]{fronczak_phase_2007}%
  \BibitemOpen
  \bibfield  {author} {\bibinfo {author} {\bibfnamefont {P.}~\bibnamefont
  {Fronczak}}, \bibinfo {author} {\bibfnamefont {A.}~\bibnamefont {Fronczak}},
  \ and\ \bibinfo {author} {\bibfnamefont {J.~A.}\ \bibnamefont {Hołyst}},\
  }\href {\doibase 10.1140/epjb/e2007-00270-8} {\bibfield  {journal} {\bibinfo
  {journal} {The European Physical Journal B}\ }\textbf {\bibinfo {volume}
  {59}},\ \bibinfo {pages} {133} (\bibinfo {year} {2007})}\BibitemShut
  {NoStop}%
\bibitem [{\citenamefont {Jeong}\ \emph {et~al.}(2007)\citenamefont {Jeong},
  \citenamefont {Choi},\ and\ \citenamefont {Park}}]{jeong_construction_2007}%
  \BibitemOpen
  \bibfield  {author} {\bibinfo {author} {\bibfnamefont {D.}~\bibnamefont
  {Jeong}}, \bibinfo {author} {\bibfnamefont {M.~Y.}\ \bibnamefont {Choi}}, \
  and\ \bibinfo {author} {\bibfnamefont {H.}~\bibnamefont {Park}},\ }\href
  {\doibase 10.1088/1751-8113/40/32/001} {\bibfield  {journal} {\bibinfo
  {journal} {Journal of Physics A: Mathematical and Theoretical}\ }\textbf
  {\bibinfo {volume} {40}},\ \bibinfo {pages} {9723} (\bibinfo {year}
  {2007})}\BibitemShut {NoStop}%
\bibitem [{\citenamefont {Foster}\ \emph {et~al.}(2010)\citenamefont {Foster},
  \citenamefont {Foster}, \citenamefont {Paczuski},\ and\ \citenamefont
  {Grassberger}}]{foster_communities_2010}%
  \BibitemOpen
  \bibfield  {author} {\bibinfo {author} {\bibfnamefont {D.}~\bibnamefont
  {Foster}}, \bibinfo {author} {\bibfnamefont {J.}~\bibnamefont {Foster}},
  \bibinfo {author} {\bibfnamefont {M.}~\bibnamefont {Paczuski}}, \ and\
  \bibinfo {author} {\bibfnamefont {P.}~\bibnamefont {Grassberger}},\ }\href
  {\doibase 10.1103/PhysRevE.81.046115} {\bibfield  {journal} {\bibinfo
  {journal} {Physical Review E}\ }\textbf {\bibinfo {volume} {81}},\ \bibinfo
  {pages} {046115} (\bibinfo {year} {2010})}\BibitemShut {NoStop}%
\bibitem [{\citenamefont {Nowicki}\ and\ \citenamefont
  {Snijders}(2001)}]{nowicki_estimation_2001}%
  \BibitemOpen
  \bibfield  {author} {\bibinfo {author} {\bibfnamefont {K.}~\bibnamefont
  {Nowicki}}\ and\ \bibinfo {author} {\bibfnamefont {T.~A.~B.}\ \bibnamefont
  {Snijders}},\ }\href {\doibase 10.1198/016214501753208735} {\bibfield
  {journal} {\bibinfo  {journal} {Journal of the American Statistical
  Association}\ }\textbf {\bibinfo {volume} {96}},\ \bibinfo {pages} {1077}
  (\bibinfo {year} {2001})}\BibitemShut {NoStop}%
\bibitem [{\citenamefont {Bender}(1974)}]{bender_asymptotic_1974}%
  \BibitemOpen
  \bibfield  {author} {\bibinfo {author} {\bibfnamefont {E.~A.}\ \bibnamefont
  {Bender}},\ }\href {\doibase 10.1016/0012-365X(74)90118-6} {\bibfield
  {journal} {\bibinfo  {journal} {Discrete Mathematics}\ }\textbf {\bibinfo
  {volume} {10}},\ \bibinfo {pages} {217} (\bibinfo {year} {1974})}\BibitemShut
  {NoStop}%
\bibitem [{\citenamefont {Bender}\ and\ \citenamefont
  {Butler}(1978)}]{bender_asymptotic_1978}%
  \BibitemOpen
  \bibfield  {author} {\bibinfo {author} {\bibfnamefont {E.}~\bibnamefont
  {Bender}}\ and\ \bibinfo {author} {\bibfnamefont {J.}~\bibnamefont
  {Butler}},\ }\href {\doibase 10.1109/TC.1978.1675021} {\bibfield  {journal}
  {\bibinfo  {journal} {Computers, {IEEE} Transactions on}\ }\textbf {\bibinfo
  {volume} {C-27}},\ \bibinfo {pages} {1180} (\bibinfo {year}
  {1978})}\BibitemShut {NoStop}%
\bibitem [{\citenamefont {Wormald}(1999)}]{wormald_models_1999}%
  \BibitemOpen
  \bibfield  {author} {\bibinfo {author} {\bibfnamefont {N.}~\bibnamefont
  {Wormald}},\ }\href@noop {} {\bibfield  {journal} {\bibinfo  {journal}
  {London Mathematical Society Lecture Note Series}\ ,\ \bibinfo {pages}
  {239–298}} (\bibinfo {year} {1999})}\BibitemShut {NoStop}%
\bibitem [{\citenamefont {Chung}\ and\ \citenamefont
  {Lu}(2002{\natexlab{a}})}]{chung_average_2002}%
  \BibitemOpen
  \bibfield  {author} {\bibinfo {author} {\bibfnamefont {F.}~\bibnamefont
  {Chung}}\ and\ \bibinfo {author} {\bibfnamefont {L.}~\bibnamefont {Lu}},\
  }\href {\doibase 10.1073/pnas.252631999} {\bibfield  {journal} {\bibinfo
  {journal} {Proceedings of the National Academy of Sciences}\ }\textbf
  {\bibinfo {volume} {99}},\ \bibinfo {pages} {15879 } (\bibinfo {year}
  {2002}{\natexlab{a}})}\BibitemShut {NoStop}%
\bibitem [{\citenamefont {Chung}\ and\ \citenamefont
  {Lu}(2002{\natexlab{b}})}]{chung_connected_2002}%
  \BibitemOpen
  \bibfield  {author} {\bibinfo {author} {\bibfnamefont {F.}~\bibnamefont
  {Chung}}\ and\ \bibinfo {author} {\bibfnamefont {L.}~\bibnamefont {Lu}},\
  }\href {\doibase 10.1007/PL00012580} {\bibfield  {journal} {\bibinfo
  {journal} {Annals of Combinatorics}\ }\textbf {\bibinfo {volume} {6}},\
  \bibinfo {pages} {125} (\bibinfo {year} {2002}{\natexlab{b}})}\BibitemShut
  {NoStop}%
\bibitem [{\citenamefont {Cover}\ and\ \citenamefont
  {Thomas}(1991)}]{cover_elements_1991}%
  \BibitemOpen
  \bibfield  {author} {\bibinfo {author} {\bibfnamefont {T.~M.}\ \bibnamefont
  {Cover}}\ and\ \bibinfo {author} {\bibfnamefont {J.~A.}\ \bibnamefont
  {Thomas}},\ }\href@noop {} {\emph {\bibinfo {title} {Elements of Information
  Theory}}},\ \bibinfo {edition} {99th}\ ed.\ (\bibinfo  {publisher}
  {{Wiley-Interscience}},\ \bibinfo {year} {1991})\BibitemShut {NoStop}%
\bibitem [{Note1()}]{Note1}%
  \BibitemOpen
  \bibinfo {note} {Because of the similarity of Eq.~\ref {eq:fermi} with the
  Fermi-Dirac distribution in quantum mechanics, as well as the analogy of the
  simple graph restriction with the Pauli exclusion principle, this type of
  ensemble is sometimes called ``fermionic'', and conversely the multigraph
  ensemble of Sec.~\ref {sec:multigraphs} is called ``bosonic''~\cite
  {park_statistical_2004}. Note however that the ``classical'' limit
  represented by Eq.~\ref {eq:classical_limit} is still insufficient to make
  these ensembles equivalent. This is only achieved by the stronger sparsity
  condition given by Eq.~\ref {eq:sparse_limit}.}\BibitemShut {Stop}%
\bibitem [{\citenamefont {Boguñá}\ \emph {et~al.}(2004)\citenamefont
  {Boguñá}, \citenamefont {{Pastor-Satorras}},\ and\ \citenamefont
  {Vespignani}}]{boguna_cut-offs_2004}%
  \BibitemOpen
  \bibfield  {author} {\bibinfo {author} {\bibfnamefont {M.}~\bibnamefont
  {Boguñá}}, \bibinfo {author} {\bibfnamefont {R.}~\bibnamefont
  {{Pastor-Satorras}}}, \ and\ \bibinfo {author} {\bibfnamefont
  {A.}~\bibnamefont {Vespignani}},\ }\href {\doibase
  10.1140/epjb/e2004-00038-8} {\bibfield  {journal} {\bibinfo  {journal} {The
  European Physical Journal B - Condensed Matter}\ }\textbf {\bibinfo {volume}
  {38}},\ \bibinfo {pages} {205} (\bibinfo {year} {2004})}\BibitemShut
  {NoStop}%
\bibitem [{\citenamefont {Park}\ and\ \citenamefont
  {Newman}(2003)}]{park_origin_2003}%
  \BibitemOpen
  \bibfield  {author} {\bibinfo {author} {\bibfnamefont {J.}~\bibnamefont
  {Park}}\ and\ \bibinfo {author} {\bibfnamefont {M.~E.~J.}\ \bibnamefont
  {Newman}},\ }\href {\doibase 10.1103/PhysRevE.68.026112} {\bibfield
  {journal} {\bibinfo  {journal} {Physical Review E}\ }\textbf {\bibinfo
  {volume} {68}},\ \bibinfo {pages} {026112} (\bibinfo {year}
  {2003})}\BibitemShut {NoStop}%
\bibitem [{\citenamefont {Johnson}\ \emph {et~al.}(2010)\citenamefont
  {Johnson}, \citenamefont {Torres}, \citenamefont {Marro},\ and\ \citenamefont
  {Muñoz}}]{johnson_entropic_2010}%
  \BibitemOpen
  \bibfield  {author} {\bibinfo {author} {\bibfnamefont {S.}~\bibnamefont
  {Johnson}}, \bibinfo {author} {\bibfnamefont {J.~J.}\ \bibnamefont {Torres}},
  \bibinfo {author} {\bibfnamefont {J.}~\bibnamefont {Marro}}, \ and\ \bibinfo
  {author} {\bibfnamefont {M.~A.}\ \bibnamefont {Muñoz}},\ }\href {\doibase
  10.1103/PhysRevLett.104.108702} {\bibfield  {journal} {\bibinfo  {journal}
  {Physical Review Letters}\ }\textbf {\bibinfo {volume} {104}},\ \bibinfo
  {pages} {108702} (\bibinfo {year} {2010})}\BibitemShut {NoStop}%
\bibitem [{\citenamefont {{McKay}}\ and\ \citenamefont
  {Wormald}(1990)}]{mckay_asymptotic_1990}%
  \BibitemOpen
  \bibfield  {author} {\bibinfo {author} {\bibfnamefont {B.}~\bibnamefont
  {{McKay}}}\ and\ \bibinfo {author} {\bibfnamefont {N.}~\bibnamefont
  {Wormald}},\ }\href@noop {} {\bibfield  {journal} {\bibinfo  {journal}
  {European J. Combin}\ }\textbf {\bibinfo {volume} {11}},\ \bibinfo {pages}
  {565–580} (\bibinfo {year} {1990})}\BibitemShut {NoStop}%
\bibitem [{Note2()}]{Note2}%
  \BibitemOpen
  \bibinfo {note} {Note that not all degree sequences are allowed in the first
  place, since they must be \protect \emph {graphical}~\cite
  {erdos_graphs_1960,del_genio_all_2011}. Imposing a block structure further
  complicates things, since the graphical condition needs to be generalized to
  blockmodels. We will not pursue this here, as we consider only the
  sufficiently sparse situation, where this issue can be
  neglected.}\BibitemShut {Stop}%
\bibitem [{\citenamefont {Bollobás}(1980)}]{bollobas_probabilistic_1980}%
  \BibitemOpen
  \bibfield  {author} {\bibinfo {author} {\bibfnamefont {B.}~\bibnamefont
  {Bollobás}},\ }\href@noop {} {\bibfield  {journal} {\bibinfo  {journal}
  {Eur. J. Comb.}\ }\textbf {\bibinfo {volume} {1}},\ \bibinfo {pages} {311}
  (\bibinfo {year} {1980})}\BibitemShut {NoStop}%
\bibitem [{\citenamefont {Newman}\ \emph {et~al.}(2001)\citenamefont {Newman},
  \citenamefont {Strogatz},\ and\ \citenamefont {Watts}}]{newman_random_2001}%
  \BibitemOpen
  \bibfield  {author} {\bibinfo {author} {\bibfnamefont {M.~E.~J.}\
  \bibnamefont {Newman}}, \bibinfo {author} {\bibfnamefont {S.~H.}\
  \bibnamefont {Strogatz}}, \ and\ \bibinfo {author} {\bibfnamefont {D.~J.}\
  \bibnamefont {Watts}},\ }\href {\doibase 10.1103/PhysRevE.64.026118}
  {\bibfield  {journal} {\bibinfo  {journal} {Physical Review E}\ }\textbf
  {\bibinfo {volume} {64}},\ \bibinfo {pages} {026118} (\bibinfo {year}
  {2001})}\BibitemShut {NoStop}%
\bibitem [{\citenamefont {Newman}(2003{\natexlab{b}})}]{newman_structure_2003}%
  \BibitemOpen
  \bibfield  {author} {\bibinfo {author} {\bibfnamefont {M.~E.~J.}\
  \bibnamefont {Newman}},\ }\href@noop {} {\bibfield  {journal} {\bibinfo
  {journal} {{SIAM} Review}\ }\textbf {\bibinfo {volume} {45}},\ \bibinfo
  {pages} {167} (\bibinfo {year} {2003}{\natexlab{b}})}\BibitemShut {NoStop}%
\bibitem [{\citenamefont {King}(2004)}]{king_comment_2004}%
  \BibitemOpen
  \bibfield  {author} {\bibinfo {author} {\bibfnamefont {O.~D.}\ \bibnamefont
  {King}},\ }\href {\doibase 10.1103/PhysRevE.70.058101} {\bibfield  {journal}
  {\bibinfo  {journal} {Physical Review E}\ }\textbf {\bibinfo {volume} {70}},\
  \bibinfo {pages} {058101} (\bibinfo {year} {2004})}\BibitemShut {NoStop}%
\bibitem [{Note3()}]{Note3}%
  \BibitemOpen
  \bibinfo {note} {The difference between ensembles with ``hard'' and ``soft''
  degree constraints is analyzed in detail in~\cite {anand_gibbs_2010} for the
  case without block structures.}\BibitemShut {Stop}%
\bibitem [{\citenamefont {Janson}(2009)}]{janson_probability_2009}%
  \BibitemOpen
  \bibfield  {author} {\bibinfo {author} {\bibfnamefont {S.}~\bibnamefont
  {Janson}},\ }\href {\doibase 10.1017/S0963548308009644} {\bibfield  {journal}
  {\bibinfo  {journal} {Combinatorics, Probability and Computing}\ }\textbf
  {\bibinfo {volume} {18}},\ \bibinfo {pages} {205} (\bibinfo {year}
  {2009})}\BibitemShut {NoStop}%
\bibitem [{\citenamefont {Wormald}(1981)}]{wormald_asymptotic_1981}%
  \BibitemOpen
  \bibfield  {author} {\bibinfo {author} {\bibfnamefont {N.}~\bibnamefont
  {Wormald}},\ }\href {\doibase 10.1016/S0095-8956(81)80022-6} {\bibfield
  {journal} {\bibinfo  {journal} {Journal of Combinatorial Theory, Series B}\
  }\textbf {\bibinfo {volume} {31}},\ \bibinfo {pages} {168} (\bibinfo {year}
  {1981})}\BibitemShut {NoStop}%
\bibitem [{\citenamefont {Holland}\ and\ \citenamefont
  {Leinhardt}(1981)}]{holland_exponential_1981}%
  \BibitemOpen
  \bibfield  {author} {\bibinfo {author} {\bibfnamefont {P.~W.}\ \bibnamefont
  {Holland}}\ and\ \bibinfo {author} {\bibfnamefont {S.}~\bibnamefont
  {Leinhardt}},\ }\href {\doibase 10.2307/2287037} {\bibfield  {journal}
  {\bibinfo  {journal} {Journal of the American Statistical Association}\
  }\textbf {\bibinfo {volume} {76}},\ \bibinfo {pages} {33} (\bibinfo {year}
  {1981})}\BibitemShut {NoStop}%
\bibitem [{\citenamefont {Wasserman}\ and\ \citenamefont
  {Pattison}(1996)}]{wasserman_logit_1996}%
  \BibitemOpen
  \bibfield  {author} {\bibinfo {author} {\bibfnamefont {S.}~\bibnamefont
  {Wasserman}}\ and\ \bibinfo {author} {\bibfnamefont {P.}~\bibnamefont
  {Pattison}},\ }\href {\doibase 10.1007/BF02294547} {\bibfield  {journal}
  {\bibinfo  {journal} {Psychometrika}\ }\textbf {\bibinfo {volume} {61}},\
  \bibinfo {pages} {401} (\bibinfo {year} {1996})}\BibitemShut {NoStop}%
\bibitem [{\citenamefont {Zhao}\ \emph {et~al.}(2011)\citenamefont {Zhao},
  \citenamefont {Levina},\ and\ \citenamefont {Zhu}}]{zhao_consistency_2011}%
  \BibitemOpen
  \bibfield  {author} {\bibinfo {author} {\bibfnamefont {Y.}~\bibnamefont
  {Zhao}}, \bibinfo {author} {\bibfnamefont {E.}~\bibnamefont {Levina}}, \ and\
  \bibinfo {author} {\bibfnamefont {J.}~\bibnamefont {Zhu}},\ }\href
  {http://arxiv.org/abs/1110.3854} {\bibfield  {journal} {\bibinfo  {journal}
  {{arXiv:1110.3854}}\ } (\bibinfo {year} {2011})}\BibitemShut {NoStop}%
\bibitem [{Note4()}]{Note4}%
  \BibitemOpen
  \bibinfo {note} {We could easily use any of the other entropy expressions
  derived previously, to accommodate the diverse variants of the ensemble,
  which could be directed, mutltigraphs, etc. However, the use of the
  expressions derived for the ``hard'' degree constraints have a more limited
  validity, since it assumes stronger sparsity conditions. We focus therefore
  on ensembles with soft degree constraints, since they are more generally
  applicable.}\BibitemShut {Stop}%
\bibitem [{Note5()}]{Note5}%
  \BibitemOpen
  \bibinfo {note} {This simple method can be very inefficient in certain cases,
  specially if the network is very large, since one may always finish in local
  maxima which are far away from the global optimum. We have also used the
  better variant know as the Kernighan-Lin algorithm~\cite
  {kernighan_efficient_1970}, adapted to the blockmodel problem in~\cite
  {karrer_stochastic_2011}, which can escape such local solutions. However, for
  the simple examples considered here, we found almost no difference in the
  results.}\BibitemShut {Stop}%
\bibitem [{\citenamefont {Metropolis}\ \emph {et~al.}(1953)\citenamefont
  {Metropolis}, \citenamefont {Rosenbluth}, \citenamefont {Rosenbluth},
  \citenamefont {Teller},\ and\ \citenamefont
  {Teller}}]{metropolis_equation_1953}%
  \BibitemOpen
  \bibfield  {author} {\bibinfo {author} {\bibfnamefont {N.}~\bibnamefont
  {Metropolis}}, \bibinfo {author} {\bibfnamefont {A.~W.}\ \bibnamefont
  {Rosenbluth}}, \bibinfo {author} {\bibfnamefont {M.~N.}\ \bibnamefont
  {Rosenbluth}}, \bibinfo {author} {\bibfnamefont {A.~H.}\ \bibnamefont
  {Teller}}, \ and\ \bibinfo {author} {\bibfnamefont {E.}~\bibnamefont
  {Teller}},\ }\href {\doibase 10.1063/1.1699114} {\bibfield  {journal}
  {\bibinfo  {journal} {The Journal of Chemical Physics}\ }\textbf {\bibinfo
  {volume} {21}},\ \bibinfo {pages} {1087} (\bibinfo {year}
  {1953})}\BibitemShut {NoStop}%
\bibitem [{\citenamefont {Hastings}(1970)}]{hastings_monte_1970}%
  \BibitemOpen
  \bibfield  {author} {\bibinfo {author} {\bibfnamefont {W.~K.}\ \bibnamefont
  {Hastings}},\ }\href {\doibase 10.1093/biomet/57.1.97} {\bibfield  {journal}
  {\bibinfo  {journal} {Biometrika}\ }\textbf {\bibinfo {volume} {57}},\
  \bibinfo {pages} {97 } (\bibinfo {year} {1970})}\BibitemShut {NoStop}%
\bibitem [{\citenamefont {Taylor}(1981)}]{mcavaney_constrained_1981}%
  \BibitemOpen
  \bibfield  {author} {\bibinfo {author} {\bibfnamefont {R.}~\bibnamefont
  {Taylor}},\ }in\ \href
  {http://www.springerlink.com/content/3777249743r61112/} {\emph {\bibinfo
  {booktitle} {Combinatorial Mathematics {VIII}}}},\ Vol.\ \bibinfo {volume}
  {884},\ \bibinfo {editor} {edited by\ \bibinfo {editor} {\bibfnamefont
  {K.~L.}\ \bibnamefont {{McAvaney}}}}\ (\bibinfo  {publisher} {Springer Berlin
  Heidelberg},\ \bibinfo {year} {1981})\ pp.\ \bibinfo {pages}
  {314--336}\BibitemShut {NoStop}%
\bibitem [{\citenamefont {Eggleton}\ and\ \citenamefont
  {Holton}(1981)}]{mcavaney_simple_1981}%
  \BibitemOpen
  \bibfield  {author} {\bibinfo {author} {\bibfnamefont {R.~B.}\ \bibnamefont
  {Eggleton}}\ and\ \bibinfo {author} {\bibfnamefont {D.~A.}\ \bibnamefont
  {Holton}},\ }in\ \href
  {http://www.springerlink.com/content/013l7u18706v604n/} {\emph {\bibinfo
  {booktitle} {Combinatorial Mathematics {VIII}}}},\ Vol.\ \bibinfo {volume}
  {884},\ \bibinfo {editor} {edited by\ \bibinfo {editor} {\bibfnamefont
  {K.~L.}\ \bibnamefont {{McAvaney}}}}\ (\bibinfo  {publisher} {Springer Berlin
  Heidelberg},\ \bibinfo {year} {1981})\ pp.\ \bibinfo {pages}
  {155--172}\BibitemShut {NoStop}%
\bibitem [{\citenamefont {Coolen}\ \emph {et~al.}(2009)\citenamefont {Coolen},
  \citenamefont {Martino},\ and\ \citenamefont
  {Annibale}}]{coolen_constrained_2009}%
  \BibitemOpen
  \bibfield  {author} {\bibinfo {author} {\bibfnamefont {A.~C.~C.}\
  \bibnamefont {Coolen}}, \bibinfo {author} {\bibfnamefont {A.}~\bibnamefont
  {Martino}}, \ and\ \bibinfo {author} {\bibfnamefont {A.}~\bibnamefont
  {Annibale}},\ }\href {\doibase 10.1007/s10955-009-9821-2} {\bibfield
  {journal} {\bibinfo  {journal} {Journal of Statistical Physics}\ }\textbf
  {\bibinfo {volume} {136}},\ \bibinfo {pages} {1035} (\bibinfo {year}
  {2009})}\BibitemShut {NoStop}%
\bibitem [{Note6()}]{Note6}%
  \BibitemOpen
  \bibinfo {note} {This algorithm actually generates samples from the \protect
  \emph {canonical} ensemble, since it allows for fluctuations in the numbers
  $e_{rs}$. However, as mentioned in Sec.~\ref {sec:def}, this ensemble is
  equivalent to the microcanonical version for sufficiently large
  samples.}\BibitemShut {Stop}%
\bibitem [{Note7()}]{Note7}%
  \BibitemOpen
  \bibinfo {note} {An alternative which circumvents this problem is the
  so-called Maximum A Posteriori (MAP) approach, which uses parameter \protect
  \emph {distributions}, instead of a single set of parameters when maximizing
  the log-likelihood. Instead of the log-likelihood increasing monotonically
  with $B$, the parameter distributions become broader instead. This approach
  has been applied to the degree-corrected stochastic blockmodel in~\cite
  {reichardt_interplay_2011}, using belief propagation. This method, however,
  has the disadvantage of being numerically less efficient for large
  networks.}\BibitemShut {Stop}%
\bibitem [{\citenamefont {Treves}\ and\ \citenamefont
  {Panzeri}(1995)}]{treves_upward_1995}%
  \BibitemOpen
  \bibfield  {author} {\bibinfo {author} {\bibfnamefont {A.}~\bibnamefont
  {Treves}}\ and\ \bibinfo {author} {\bibfnamefont {S.}~\bibnamefont
  {Panzeri}},\ }\href {\doibase i: 10.1162/neco.1995.7.2.399</p>} {\bibfield
  {journal} {\bibinfo  {journal} {Neural Computation}\ }\textbf {\bibinfo
  {volume} {7}},\ \bibinfo {pages} {399} (\bibinfo {year} {1995})}\BibitemShut
  {NoStop}%
\bibitem [{Note8()}]{Note8}%
  \BibitemOpen
  \bibinfo {note} {We note that Eq.~\ref {eq:bias} should be understood only as
  a rule of thumb which gives a \protect \emph {lower bound} on the bias of
  $\protect \mathcal {L}$, since it is obtained only from the first term of
  Eq.~\ref {eq:L}, and assumes that the number of blocks in each partition
  fluctuates independently, which is not likely to hold in general since the
  block partition is a result of an optimization algorithm.}\BibitemShut
  {Stop}%
\bibitem [{\citenamefont {Bianconi}\ \emph {et~al.}(2009)\citenamefont
  {Bianconi}, \citenamefont {Pin},\ and\ \citenamefont
  {Marsili}}]{bianconi_assessing_2009}%
  \BibitemOpen
  \bibfield  {author} {\bibinfo {author} {\bibfnamefont {G.}~\bibnamefont
  {Bianconi}}, \bibinfo {author} {\bibfnamefont {P.}~\bibnamefont {Pin}}, \
  and\ \bibinfo {author} {\bibfnamefont {M.}~\bibnamefont {Marsili}},\ }\href
  {\doibase 10.1073/pnas.0811511106} {\bibfield  {journal} {\bibinfo  {journal}
  {Proceedings of the National Academy of Sciences}\ }\textbf {\bibinfo
  {volume} {106}},\ \bibinfo {pages} {11433 } (\bibinfo {year}
  {2009})}\BibitemShut {NoStop}%
\bibitem [{\citenamefont {Lancichinetti}\ \emph {et~al.}(2010)\citenamefont
  {Lancichinetti}, \citenamefont {Radicchi},\ and\ \citenamefont
  {Ramasco}}]{lancichinetti_statistical_2010}%
  \BibitemOpen
  \bibfield  {author} {\bibinfo {author} {\bibfnamefont {A.}~\bibnamefont
  {Lancichinetti}}, \bibinfo {author} {\bibfnamefont {F.}~\bibnamefont
  {Radicchi}}, \ and\ \bibinfo {author} {\bibfnamefont {J.~J.}\ \bibnamefont
  {Ramasco}},\ }\href {\doibase 10.1103/PhysRevE.81.046110} {\bibfield
  {journal} {\bibinfo  {journal} {Physical Review E}\ }\textbf {\bibinfo
  {volume} {81}},\ \bibinfo {pages} {046110} (\bibinfo {year}
  {2010})}\BibitemShut {NoStop}%
\bibitem [{\citenamefont {Radicchi}\ \emph {et~al.}(2010)\citenamefont
  {Radicchi}, \citenamefont {Lancichinetti},\ and\ \citenamefont
  {Ramasco}}]{radicchi_combinatorial_2010}%
  \BibitemOpen
  \bibfield  {author} {\bibinfo {author} {\bibfnamefont {F.}~\bibnamefont
  {Radicchi}}, \bibinfo {author} {\bibfnamefont {A.}~\bibnamefont
  {Lancichinetti}}, \ and\ \bibinfo {author} {\bibfnamefont {J.~J.}\
  \bibnamefont {Ramasco}},\ }\href {\doibase 10.1103/PhysRevE.82.026102}
  {\bibfield  {journal} {\bibinfo  {journal} {Physical Review E}\ }\textbf
  {\bibinfo {volume} {82}},\ \bibinfo {pages} {026102} (\bibinfo {year}
  {2010})}\BibitemShut {NoStop}%
\bibitem [{\citenamefont {Lancichinetti}\ \emph {et~al.}(2011)\citenamefont
  {Lancichinetti}, \citenamefont {Radicchi}, \citenamefont {Ramasco},\ and\
  \citenamefont {Fortunato}}]{lancichinetti_finding_2011}%
  \BibitemOpen
  \bibfield  {author} {\bibinfo {author} {\bibfnamefont {A.}~\bibnamefont
  {Lancichinetti}}, \bibinfo {author} {\bibfnamefont {F.}~\bibnamefont
  {Radicchi}}, \bibinfo {author} {\bibfnamefont {J.~J.}\ \bibnamefont
  {Ramasco}}, \ and\ \bibinfo {author} {\bibfnamefont {S.}~\bibnamefont
  {Fortunato}},\ }\href {\doibase 10.1371/journal.pone.0018961} {\bibfield
  {journal} {\bibinfo  {journal} {{PLoS} {ONE}}\ }\textbf {\bibinfo {volume}
  {6}},\ \bibinfo {pages} {e18961} (\bibinfo {year} {2011})}\BibitemShut
  {NoStop}%
\bibitem [{\citenamefont {Erdős}\ and\ \citenamefont
  {Gallai}(1960)}]{erdos_graphs_1960}%
  \BibitemOpen
  \bibfield  {author} {\bibinfo {author} {\bibfnamefont {P.}~\bibnamefont
  {Erdős}}\ and\ \bibinfo {author} {\bibfnamefont {T.}~\bibnamefont
  {Gallai}},\ }\href@noop {} {\bibfield  {journal} {\bibinfo  {journal} {Mat.
  Lapok.}\ }\textbf {\bibinfo {volume} {11}},\ \bibinfo {pages} {264} (\bibinfo
  {year} {1960})}\BibitemShut {NoStop}%
\bibitem [{\citenamefont {Del~Genio}\ \emph {et~al.}(2011)\citenamefont
  {Del~Genio}, \citenamefont {Gross},\ and\ \citenamefont
  {Bassler}}]{del_genio_all_2011}%
  \BibitemOpen
  \bibfield  {author} {\bibinfo {author} {\bibfnamefont {C.~I.}\ \bibnamefont
  {Del~Genio}}, \bibinfo {author} {\bibfnamefont {T.}~\bibnamefont {Gross}}, \
  and\ \bibinfo {author} {\bibfnamefont {K.~E.}\ \bibnamefont {Bassler}},\
  }\href {\doibase 10.1103/PhysRevLett.107.178701} {\bibfield  {journal}
  {\bibinfo  {journal} {Physical Review Letters}\ }\textbf {\bibinfo {volume}
  {107}},\ \bibinfo {pages} {178701} (\bibinfo {year} {2011})}\BibitemShut
  {NoStop}%
\bibitem [{\citenamefont {Kernighan}\ and\ \citenamefont
  {Lin}(1970)}]{kernighan_efficient_1970}%
  \BibitemOpen
  \bibfield  {author} {\bibinfo {author} {\bibfnamefont {B.}~\bibnamefont
  {Kernighan}}\ and\ \bibinfo {author} {\bibfnamefont {S.}~\bibnamefont
  {Lin}},\ }\href@noop {} {\bibfield  {journal} {\bibinfo  {journal} {Bell
  System Technical Journal}\ }\textbf {\bibinfo {volume} {49}},\ \bibinfo
  {pages} {291–307} (\bibinfo {year} {1970})}\BibitemShut {NoStop}%
\end{thebibliography}%
\end{document}